\documentclass[a4paper, twocolumn,accepted=2025-03-31]{quantumarticle}
\pdfoutput=1
\usepackage{graphicx}% Include figure files
\usepackage{bm}% bold math
\usepackage{bbold}
\usepackage{physics}
\usepackage{xcolor}
\usepackage{enumitem}
\usepackage{amsmath, amssymb}
\usepackage[normalem]{ulem}
\usepackage{physics}
\usepackage{comment}
\usepackage{diagbox}
\usepackage{dsfont}
\usepackage{adjustbox}
\usepackage{array}
\usepackage{float}
\usepackage{booktabs}

\usepackage[colorlinks=true]{hyperref}
\hypersetup{citecolor = blue, linkcolor=blue, urlcolor=blue}

\newtheorem{fact}{Fact}[section]
\begin{document}

\title{Charge and Spin Sharpening Transitions on Dynamical {Quantum} Trees}

\author{Xiaozhou Feng}
\affiliation{Department of Physics, The University of Texas at Austin, Austin, TX 78712, USA}

\author{Nadezhda Fishchenko}
\affiliation{Physics Department, Princeton University, Princeton, New Jersey 08540, USA}

\author{Sarang Gopalakrishnan}
\affiliation{Department of Electrical and Computer Engineering, 
Princeton University, Princeton, New Jersey 08540, USA}

\author{Matteo Ippoliti}
\affiliation{Department of Physics, The University of Texas at Austin, Austin, TX 78712, USA}

\begin{abstract}

The dynamics of monitored systems can exhibit a measurement-induced phase transition (MIPT) between mixed and pure phases, tuned by the measurement rate. When the dynamics obeys a continuous symmetry, the mixed phase further splits into a fuzzy phase and a sharp phase based on the scaling of fluctuations of the symmetry charge. While the sharpening transition for Abelian symmetries is well understood analytically, no such understanding exists for the non-Abelian case. In this work, building on a recent analytical solution of the MIPT on tree-like circuit architectures (where qubits are repatedly added or removed from the system in a recursive pattern), we study purification and sharpening transitions in monitored dynamical quantum trees obeying $U(1)$ and $SU(2)$ symmetries. The recursive structure of tree tensor networks enables powerful analytical and numerical methods to determine the phase diagrams in both cases. In the $U(1)$ case, we analytically derive a Fisher-KPP-like differential equation that allows us to locate the critical point and identify its properties. We find that the purification and sharpening transitions generically occur at distinct measurement rates. In the $SU(2)$ case, we find that the fuzzy phase is generic, and a sharp phase is possible only in the limit of maximal measurement rate. In this limit, we analytically solve the boundaries separating the fuzzy and sharp phases, and find them to be in agreement with exact numerical simulations.
\end{abstract}

\maketitle

\tableofcontents

\section{Introduction}

Monitored quantum dynamics is a class of quantum dynamical processes in which unitary evolution alternates with measurements~\cite{Skinner_2019,Li_2018,Chan_2019}. The measurements can be either weak or projective~\cite{Szyniszewski_Entanglement_2019,Szyniszewski_Universality_2020}; what matters is that the measurement outcome is recorded, and one considers the properties of the quantum state that is obtained \emph{conditional} on the measurement record. It was recently realized that monitored dynamics generically gives rise to two phases, separated by a phase transition upon tuning the measurement rate. This is called the measurement induced phase transition (MIPT) \cite{Li_Measurement-driven_2019,Choi_2020,Gullans_2020,Bao_Theory_2020,Jian_Measurement_2020,Li_Conformal_2021,Zabolo_Critical_2020,Gullans_Scalable_2020,Tang_Measurement-induced_2020,Turkeshi_Measurement-induced_2020,Fan_Self-organized_2021,Li_Statistical_2021,nahum_measurement_2021,li2021statistical,Lavasani_Measurement_2021,Ippoliti_Entanglement_2021,Lopez-Piqueres_Mean-field_2020,Noel_2022,Ippoliti_postselection_2021,Koh_Experimental_2022,Li_Cross_2022,li_entanglement_2023,feng_measurement_2023,Hoke_Measurement-induced_2023,Fava_Nonlinear_2023,jian2023measurementinduced,agrawal_entanglement_2022,learnability_Fergus,majidy_critical_2023,Ippoliti_Learnability_2024,akhtar2023measurementinduced,garratt2023probing,mcginley_postselection-free_2023}. The MIPT separates a phase (at low measurement rate) that is called the mixed or entangling phase, and one at high measurement rate that is called the pure or disentangling phase. The MIPT is invisible in local expectation values, but manifests itself in a number of information-theoretic probes. The MIPT was first characterized in terms of bipartite entanglement: starting from a pure state and running the monitored dynamics, one finds that the entanglement of a subsystem scales with its volume in the entangling phase and its area in the disentangling phase \cite{Skinner_2019,Li_2018,Chan_2019}. Entanglement is a natural probe for systems with simple Euclidean geometries, but does not extend cleanly to nonlocal geometries. A more general characterization of the MIPT is in terms of the timescale for an initially mixed state to purify \cite{Gullans_2020}. In the entangling phase, this timescale is typically exponential in system size $N$; in the disentangling phase, it is typically $O(\log(N))$ \cite{Li_Statistical_2021}. The purification perspective relates the MIPT to the physics of quantum error correcting codes: in the mixed/entangling phase, quantum information that is encoded in the initial mixed state remains encoded against the local measurements, whereas in the disentangling phase this information is rapidly collapsed by measurements \cite{Choi_2020,nahum_measurement_2021,fidkowski2021dynamical,Li_Statistical_2021}. Since its discovery the MIPT has been studied extensively in both numerical and experimental work (see Refs.~\cite{Potter_2022,Fisher_2023} for reviews). 

In local Euclidean geometries, the MIPT is analytically intractable except in the limit of strictly infinite on-site Hilbert space dimension, where it maps to percolation \cite{Skinner_2019,Bao_Theory_2020,Nahum_Quantum_2017}. Analytic progress on the MIPT for qubits (or more generally qudits) has relied on using the purification perspective in geometries that are designed to be analytically tractable, including all-to-all connected systems and tree tensor networks \cite{Gullans_2020,nahum_measurement_2021,Lopez-Piqueres_Mean-field_2020,feng_measurement_2023,vijay2020measurementdriven}. 
The purification process in dynamical tree problems can be analytically solved by relating the dynamics to a discrete version of the Fisher-KPP equation \cite{Fisher_The_1937,Derrida_Polymers_1988,nahum_measurement_2021,feng_measurement_2023}. The critical points and the scaling exponents can be solved and belong to the same universality class as a type of glass transition~\cite{Derrida_Polymers_1988,nahum_measurement_2021,feng_measurement_2023,derrida_survival_2007}. This makes the dynamical tree model one of the rare cases of MIPT in which the phase transition can be analytically understood. Even in models that cannot be analytically solved, the tree model enables us to use efficient recursive numerical methods to get reliable numerical estimates of critical points and exponents \cite{Miller_Weak-disorder_1994,Monthus_Anderson_2009,Garcia-Mata_Scaling_2017,Shi_Classical_2006,Murg_Simulating_2010,Silvi_Homogeneous_2010,Tagliacozzo_Simulation_2009,Li_Efficient_2012}.

The usual MIPT occurs in systems where the dynamics obeys no symmetries. Continuous symmetries, in particular, are thought to destabilize the critical point of the standard MIPT~\cite{FieldTheory_Fergus}. More dramatically, in dynamics obeying a $U(1)$ symmetry, the entangling phase splits into two different phases depending on whether measurements are effective at collapsing superpositions of states with different charge~\cite{agrawal_entanglement_2022}, or equivalently whether the symmetry charge can be \emph{learned} by an eavesdropper given access to the measurement data~\cite{learnability_Fergus, agrawal2023observing}. In the charge-fuzzy phase, the global conserved charge of a system of $N$ qubits is learned in $O(N)$ time, whereas in the charge-sharp phase it is learned in $O(\log N)$ time. In one dimension the $U(1)$ sharpening transition is analytically understood~\cite{FieldTheory_Fergus}, but in higher dimensions it remains an open question. Even in one dimension, extending the symmetry group to $SU(2)$ leads to a confusing phase diagram~\cite{majidy_critical_2023}, featuring two apparently distinct phases that are both ``fuzzy'' in that the charge takes longer than $O(N)$ time to learn.

In this paper we consider the MIPT in dynamical tree models with $U(1)$ or $SU(2)$ symmetry. We will specify the tree models in detail in the next section; in each case, we have chosen the simplest tree geometry that captures the relevant physics. For $U(1)$-symmetric trees, we find both a purification and a sharpening transition. We further point out that these transitions can still be analytically studied using methods similar to those applied in previous works. One interesting result is that although for systems of qubits the two critical points (purification and sharpening) are numerically very close, they become separated in systems of higher-dimensional qudits (where we can analytically solve the limit of infinite local Hilbert space dimension). For $SU(2)$-symmetric trees, the purification transition is generically absent. However, a spin-sharpening transition persists for a certain architecture of the tree (although the sharp phase only appears when every qubit is measured at each time step). By modifying methods introduced in previous work \cite{Derrida_Polymers_1988,nahum_measurement_2021,feng_measurement_2023}, we give an analytical solution of the phase boundary, which is further found to be in good agreement with exact numerical calculations on finite-sized systems of up to 128 qubits.

The rest of this paper is organized as follows. 
In Sec.~\ref{sec:keys} we briefly review MIPTs on dynamical trees and the sharpening transition, then provide a summary of our findings.
In Sec.~\ref{sec:U1} we address the dynamics of a tree model with $U(1)$ symmetry. In Section~\ref{sec:spin_sharpening} we turn to the case with $SU(2)$ symmetry, including both the analytical solution of phase boundary and numerical calculations on systems up to 128 qubits. 
We conclude in  Sec.~\ref{sec:concl} with a discussion of  potential future directions.

\section{Key concepts and summary of results}
\label{sec:keys}

\subsection{Purification of dynamical quantum trees}

Here we give a brief review about quantum tree model (see Ref.~\cite{nahum_measurement_2021,feng_measurement_2023} for more details). In this paper we consider two processes of quantum dynamics, \textit{the collapse process} and \textit{the expansion process}. In the former,we discard half of the system qubits by projective measurement within each timestep. We start with a $2^k$ qubits quantum state  which is usually taken to be maximally mixed and halve the system size at each time step, thus ending up with a single qubit after $k$ time steps. This gives rise to a tree with $k$ layers, as shown in Fig.~\ref{fig:U1_tree}. Within each node, there is an unitary gate entangling the two subtrees and a projective measurement used to remove one qubit from the tree. There can be one extra measurement with probability $p$ acting on the remaining qubit. By increasing the measurement rate of mid-circuit measurements, the system is shown to exhibit a purification transition, characterized by the purity of top qubit in the thermodynamic limit $k\to\infty$\cite{nahum_measurement_2021,feng_measurement_2023}.

On the other hand, the expansion tree starts with only a finite number of qubits (depending on the exact problem) in the maximally mixed state. Then in each time step we entangle the system with a number of qubits equal to the current size of the system in some pre-determined pure states. Thus the total system size doubles at each step, as shown by Fig.~\ref{fig:SU2_tree}. Similar to the collapse process, there are also mid-circuit measurements in the tree, which induce the phase transition between non-purifying and purifying phases. Since a maximally mixed state of an input qubit can be viewed as one half of a Bell pair state with a reference qubit, we can equivalently characterize the MIPT by focusing on the density matrices of the reference qubits alone~\cite{Gullans_Scalable_2020}. Note that the collapse process and expansion process can be mapped to each other by a time reversal collapse process as the inverse of expansion \cite{feng_measurement_2023}. However, since the initial states of ancilla qubits in the expansion process are pre-dertermined, they must be ``forced'' measurements (i.e., projectors onto predetermined outcomes, not set by the Born rule) in the corresponding collapse process. Then an expansion process where all measurements are real (i.e. Born's rule) is mapped onto a collapse process with both real and forced measurements. Unlike the collapse process with only real or forced measurements, where an analytical theory is established \cite{nahum_measurement_2021,feng_measurement_2023}, process with both real and forced measurements generally lacks an analytical solution.

\subsection{Charge sharpening in $U(1)$-symmetric monitored dynamics}
Previous work shows that the phase diagram of monitored $1$-D systems can be enriched when the dynamics obeys a $U(1)$ charge symmetry\cite{FieldTheory_Fergus,agrawal_entanglement_2022,learnability_Fergus, agrawal2023observing}.

The charge-sharpening transition has primarily been studied in the following setting. Consider a lattice where each vertex hosts both a qubit and a $d$-level system (qudit). Define the on-site charge density operator $q_i \equiv \frac{1}{2}(1 + Z_i) \otimes \mathbb{I}$, which is insensitive to the state of the qudit. The total charge is $Q \equiv \sum_i q_i$.
This lattice evolves under a quantum circuit consisting of geometrically local quantum gates that commute with $Q$, as well as single-site measurements in a fixed reference basis, such that every Kraus operator commutes with $Q$. 

If the initial state is an eigenstate of $Q$, its charge is invariant under the circuit. Otherwise---if it is a superposition or a mixture of charge states---measurements generally collapse the state to an eigenstate of $Q$ (i.e., ``sharpen the charge''). It was found that the entangling phase splits up into two distinct phases based on the time scale over which the charge sharpens \cite{agrawal_entanglement_2022}: the `fuzzy' phase, where the timescale is extensive in system size, and the `sharp' phase where the time scale is $O(\log N)$. The phase transition between them is called the charge sharpening transition. In the limit $d\to\infty$, the dynamics can be understood as a symmetric exclusion process constrained by mid-circuit measurement outcomes \cite{agrawal_entanglement_2022}. The resulting charge-sharpening transition can be shown to be in the Kosterlitz-Thouless universality class. For the case of finite $d$, the charged (qubit) and neutral (qudit) degrees of freedom do not decouple. This coupled dynamics is not analytically tractable in general, although in one dimension, field-theoretic arguments \cite{FieldTheory_Fergus} suggest that the sharpening transition remains in the Kosterlitz-Thouless class for large enough $d$. For $d = 1$, the critical points for the sharpening transition and the entanglement transition (i.e., MIPT) appear close to each other, and cannot clearly be resolved with available numerical methods \cite{agrawal_entanglement_2022}. %the available numerical methods cannot clearly resolve them for qubits. 
For $1 < d < \infty$ (or in spatial dimensions $> 1$) the problem on a Euclidean lattice is not numerically tractable. In particular, there is still some uncertainty about the nature of the phase diagram as a function of qudit dimension $d$ and measurement rate: does the sharpening transition lie inside the volume law phase for all $d$, or do they coalesce at some point? 

\subsection{Spin sharpening in $SU(2)$-symmetric monitored dynamics}
Besides the Abelian symmetry, another interesting question is the dynamics of monitored circuits enriched by non-Abelian symmetry, the simplest example being $SU(2)$. In the $SU(2)$ case, the dynamics conserves  all components of the spin vector, rather than just its $Z$ component. This imposes additional constraints on the gates and measurements: in particular, the simplest  measurements compatible with the symmetry are two-site measurements that project the spin on two sites into either the singlet or the triplet subspace. 

Previous work \cite{majidy_critical_2023} finds that monitored dynamics with $SU(2)$ symmetry of a system of qubits in $1$D seems to exhibit an entanglement transition between a phase with purification time $t \sim O(N^2)$ (at high measurement rate) and one with $t\sim O(\exp(N))$ (at low measurement rate). These two phases also seem to have different spin sharpening timescales. The system seems to exhibit $O(N^3)$ timescale in the fuzzy phase and $O(N^2)$ timescale in the sharp phase with $N$ as the system size. Although numerical work suggests the existence of this spin sharpening transition, the analytical understanding of both purification and sharpening are hard problems due to the non-Abelian nature of the $SU(2)$-symmetry. 

\subsection{Summary of models and results}

As our review of sharpening transitions indicates, many aspects of these transitions remain unsettled: in particular, their properties in spatial dimensions greater than one, the separation between the sharpening and entanglement transitions as a function of qudit dimension $d$, and almost all questions regarding the $SU(2)$ case and the case of other nonabelian symmetries. Since the dynamics of tree circuits is generally more tractable than that on Euclidean lattices, it is natural to use these architectures to shed light on sharpening transitions in general.

\begin{figure}
    \centering
    \includegraphics[width=\columnwidth]{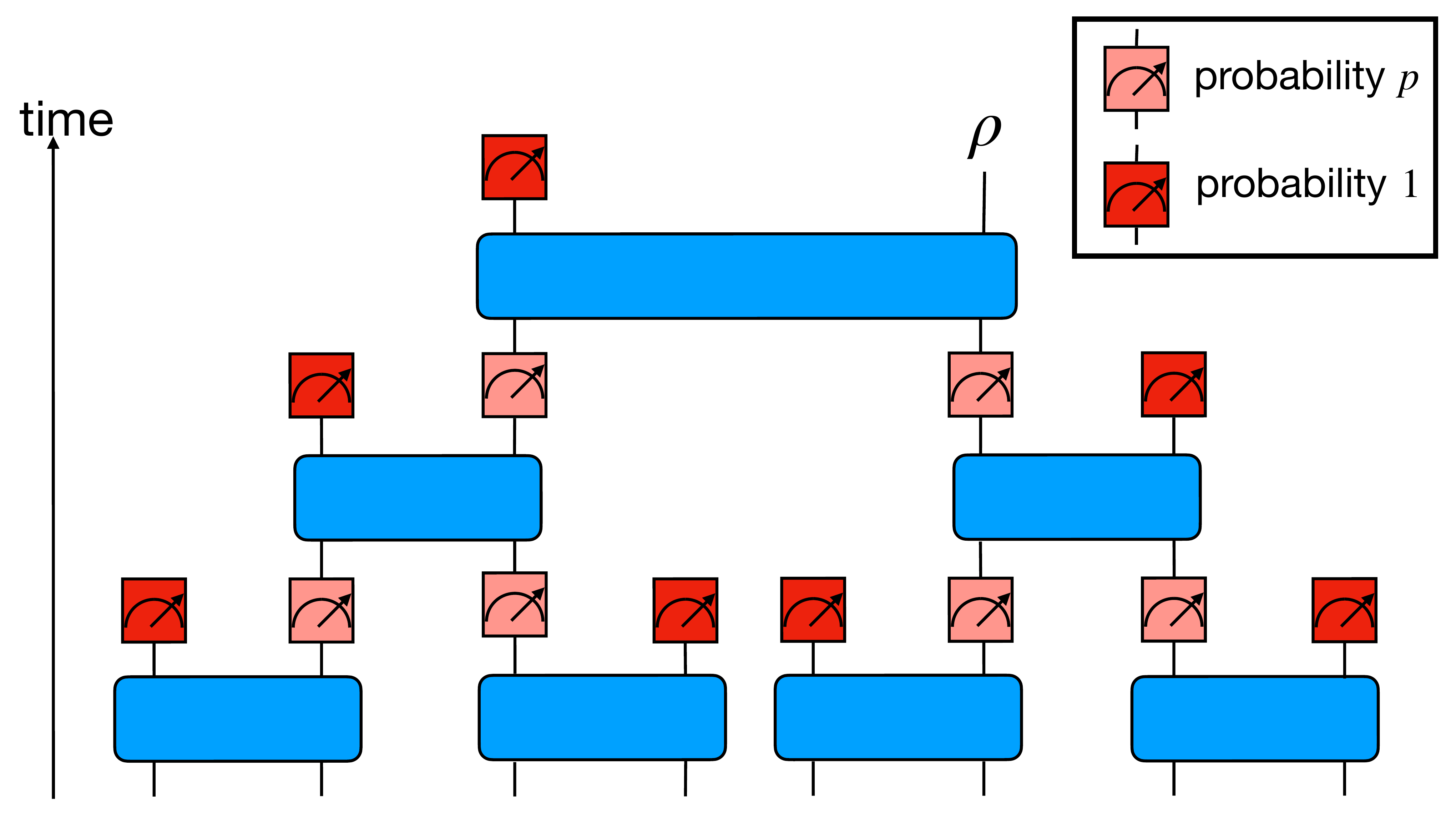}
    \caption{Illustration of our tree model with $U(1)$ symmetry. The blue blocks represent random unitary gates with the $U(1)$ symmetry. All red blocks are projective measurements which measure the site (i.e. both the charged qubit and the neutral $d$-state qudit jointly) in the computational basis. For each discrete time step, half of the system is discarded by projective measurement (darker red boxes). In the remaining half of the system, each site is measured with probability $p$ (lighter red boxes). Our goal is to obtain the density matrix of the top qubit of the tree.}
    \label{fig:U1_tree}
\end{figure}

We now introduce the models and main strategies we will use to study the $U(1)$ and $SU(2)$ sharpening transitions on dynamical trees. In the $U(1)$ case, we work with the analytically simplest tree geometry, which is the collapse tree. The model is shown in Fig.~\ref{fig:U1_tree}. As in previous work \cite{agrawal_entanglement_2022}, each site is formed by a charged qubit coupled to a neutral $d$-level system ($d\ge 1$). For a tree circuit with $k$ layers, we start with $2^k$ sites and discard half of the sites by projective measurements within each step. Each node of the tree comprises a two-site $U(1)$-symmetric entangling gate (blue boxes in Fig.~\ref{fig:U1_tree}), a projective measurement of one site (dark red box), and a projective measurement on the other site with probability $p$ (light red box). 
Since the dynamics conserves the total $z$ component of the charged qubits, we consider measurements in the computational basis, which commute with the symmetry. For all measurements in Fig.~\ref{fig:U1_tree} (either dark red or light red boxes), the charged qubit and the $d$-level system on a given site are measured simultaneously. 
All entangling gates $U$ in the tree take the form
\begin{equation}
    U =
    \begin{pmatrix}
        U^0_{d^2\times d^2}&\\&U^1_{2d^2\times 2d^2}&\\
        &&U^2_{d^2\times d^2}
    \end{pmatrix},
\end{equation}
where $U^q$ are Haar random unitary gates in the sector with total charge $q={0,1,2}$. For the case of $d=1$, $U^0$ and $U^2$ reduce to uniformly random phase factors and $U^1$ is a $2\times 2$ unitary matrix. 

At the root of the tree one has a single site, whose state captures the dynamics of the whole system. 
We are interested in the final state of this site in the ``thermodynamic limit''\footnote{It is worth noting that this is jointly a thermodynamic limit and late-time limit as the two concepts are not independent on the tree geometry.} $k\to\infty$. 
More specifically, we start with either the maximally mixed state (purification) or the pure state $\ket{\psi_0}=\ket{+}^{\otimes 2^k}$ with equal probability for all bitstrings (charge sharpening) and ask whether the final state is purified/sharpened when the total number of layers $k$ goes to infinity. Due to the recursive structure of the tree, we can view the density matrix\footnote{
In previous work about the MIPT in tree models~\cite{nahum_measurement_2021,feng_measurement_2023},  the only relevant information about the state of the top qubit is its purity. This is due to local Haar invariance of the circuit, which guarantees that the basis of the density matrix can be safely ignored. In the problem we consider here, this is no longer true due to the $U(1)$-symmetry which picks out a preferential direction (the $z$ direction in our case). So we need to keep track of not just the purity, but also of which $z$ basis state the density matrices are closest to when considering the recursive construction of the tree.
} 
for the qubit at the top of the depth-$k$ tree as the output of a map that acts on the two density matrices of depth-$(k-1)$ subtrees, as shown by Fig.~\ref{fig:U1_tree}. This allows us to write down a recursive relation for the density matrices\footnote{
More precisely, this recursive approach works only when there is no mix of ``forced'' measurements (i.e. projectors on a predetermined outcome state) and ``real'' measurements (where the outcome follows the Born rule). For the $U(1)$ case, this is guaranteed since we only have real measurements. However, this problem will arise in the $SU(2)$-symmetric case, complicating the numerical study of the problem. We discuss this in Sec.~\ref{sec:spin_sharpening}.}, 
which generally captures all the information we need to characterize purification and sharpening. 
In the limit of large qudit dimension $d \to \infty$ this tree leads to sufficiently simple recursion relations that one can explicitly solve for the sharpening transition. In addition to locating this transition, we compute distribution functions of the purity of the site at the top of the tree, identifying the sharpening transition as being in the ``glass'' universality class\footnote{
The ``glass'' universality class here refers originally to polymers in disordered environments~\cite{Derrida_Polymers_1988}, and was recently found in the context of monitored circuits on tree-like geometries~\cite{nahum_measurement_2021}. Unlike the BKT transition found in previous works on charge sharpening~\cite{agrawal_entanglement_2022}, this ``glass'' transition is second-order. It is due to the competition between the branching of the tree and the disorder (in our case represented by measurements) which can cut branches of the tree. When the disorder strength (measurement rate in our setup) is below a critical point, the branching dominates and the tree remains connected; in the other phase, the disorder dominates and the tree is disconnected giving a vanishing order parameter. When we approach the phase transition from below, the order parameter vanishes exponentially. See Ref.~\cite{nahum_measurement_2021} for more details about this universality class.
}
where this distribution is fat-tailed \cite{Derrida_Polymers_1988,nahum_measurement_2021,feng_measurement_2023}. We supplement our analytic treatment of $d = \infty$ with numerics on large trees for $d = 1, 2, 3$. For $d = 1$, the transitions coincide in this geometry for trivial reasons (this is unlike models in one spatial dimension~\cite{agrawal_entanglement_2022}); however, even at $d = 2$, the sharpening and purification transitions are clearly separate. Unexpectedly, the critical point for the sharpening transition evolves non-monotonically with $d$, decreasing from $d = 1$ to $d = 2$ before rising again at larger $d$.

%In the $U(1)$ case, we work with the analytically simplest tree geometry, which is the collapse tree. The collapse tree contains two-site gates and single-site measurements, so incorporating $U(1)$ symmetry is straightforward. In the limit of large qudit dimension $d = \infty$ this tree leads to sufficiently simple recursion relations that one can explicitly solve for the sharpening transition. In addition to locating this transition, we compute distribution functions of the purity of the site at the top of the tree, identifying the sharpening transition as being in the ``glass'' universality class where this distribution is fat-tailed \cite{Derrida_Polymers_1988,nahum_measurement_2021,feng_measurement_2023}. Different from the BKT transition in previous work \cite{agrawal_entanglement_2022}, this is a 2d phase transition. We supplement our analytic treatment of $d = \infty$ with numerics on large trees for $d = 1, 2, 3$. For $d = 1$, the transitions coincide in this geometry for trivial reasons, which is different from that in 1$d$ circuit \cite{agrawal_entanglement_2022}; however, even at $d = 2$, the sharpening and purification transitions are clearly separate. Unexpectedly, the critical point for the sharpening transition evolves non-monotonically with $d$, decreasing from $d = 1$ to $d = 2$ before rising again at larger $d$.

\begin{figure}
    \centering
    \includegraphics[width=1.0\columnwidth]{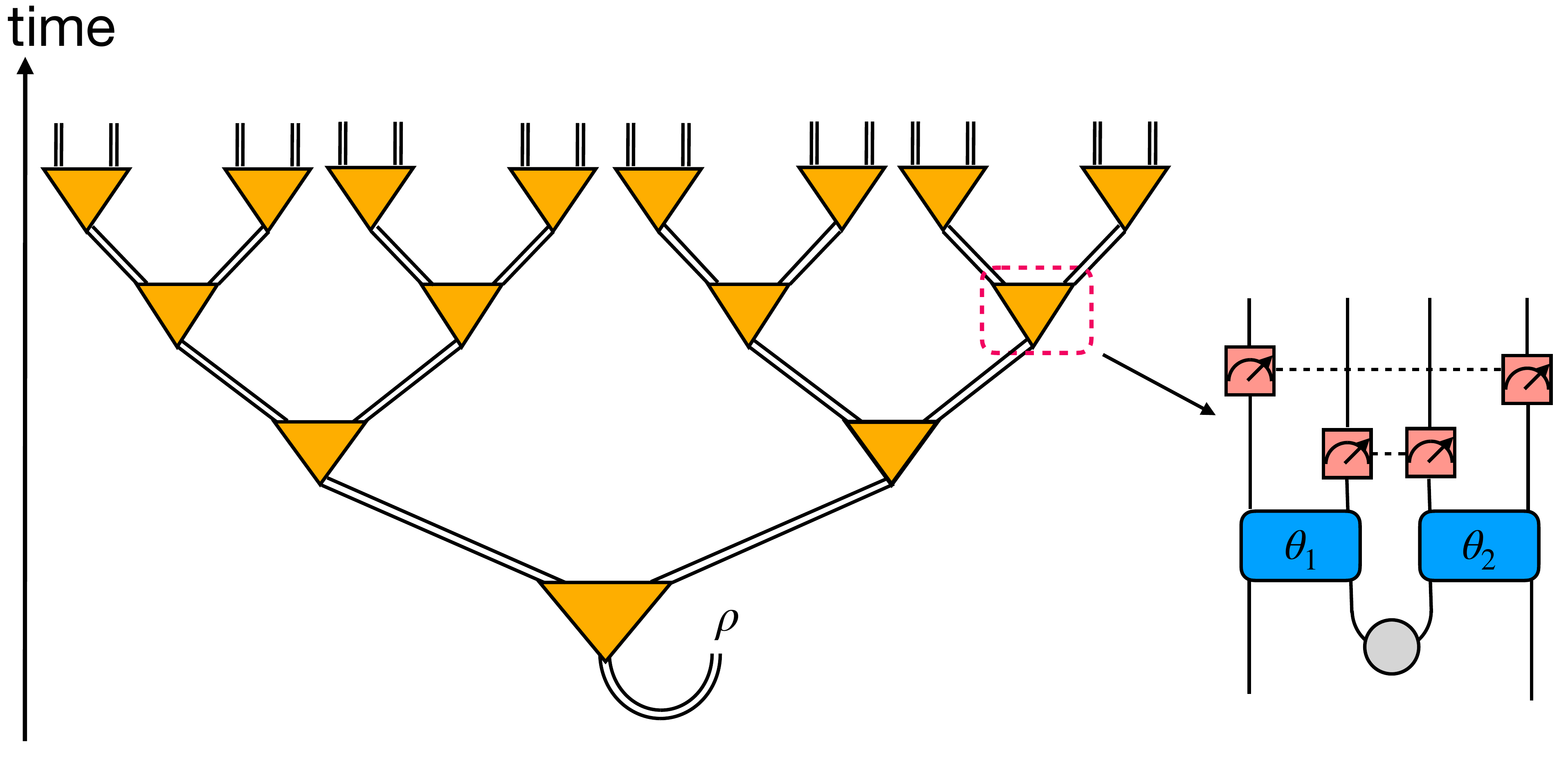}
    \caption{Schematic of our expansion tree model with $SU(2)$ symmetry. The inner structure of a node (dashed box) is shown on the right. Two new qubits in a singlet state (represented by the gray circle) are introduced in each node besides the two input qubits. The qubits are coupled by two-qubit unitary gates with $SU(2)$ symmetry, each parametrized by an angle $\theta$, in the pattern shown. The tunable parameters $\theta_1$ and $\theta_2$ are fixed throughout the dynamics. After the entangling gates, measurements of two-qubit $SU(2)$-symmetric observables $\boldsymbol{\sigma}_i\cdot \boldsymbol{\sigma}_j$ happen with probability $p$, in the `crossed' pattern shown (dashed lines denote the qubit pairs).
    Each measurement projects the qubit pair into either a singlet or a triplet state. At the initial time, the two system qubits are maximally entangled with two reference qubits forming Bell pairs. We study the reduced density matrix $\rho$ of the two reference qubits.}
    \label{fig:SU2_tree}
\end{figure}

For the $SU(2)$ case, defining a suitable tree model is more delicate. First, we note that the minimal measurement involves two sites, and only reveals if those sites are in a singlet or a triplet. Since the triplet projector is not rank-1, it does not completely decouple the measured pair of qubits. The collapse process requires rank-1 projectors to be well-defined, so the only feasible options are (1) processes involving forced singlet measurements everywhere, and (2) the expansion process. To keep the physics of the Born rule, we choose to work with the expansion process.
Our model is shown in Fig.~\ref{fig:SU2_tree}. We start with two initial qubits which form two Bell pairs with two reference qubits. These two reference qubits behave as the probe of the dynamics. At late time, we are interested in the final state of the two probe qubits conditioned on all mid-circuit measurement outcomes. During the time evolution, we repeatedly introduce ancilla qubit pairs initialized in a singlet state, thus preserving $SU(2)$ symmetry. This forms a tree model representing an expansion process. Up to an overall phase, $SU(2)$-symmetric unitary gates on two qubits take the single-parameter form \cite{majidy_critical_2023}
\begin{equation}
    U = e^{-i\theta}P_s+e^{i\theta}P_t,
\end{equation}
with $P_s$ and $P_t$ the singlet and triplet projectors. The angles $\theta_1$ and $\theta_2$ for the two unitary gates in each node are fixed across all nodes in the dynamics, and serve as tunable parameters for our dynamics.  

Besides the entangling gates, we allow mid-circuit measurements in each node. The $SU(2)$ symmetry rules out single-qubit measurements; we choose to measure the total spin on pairs of qubits, in the ``crossed'' pattern\footnote{The other possible pairing would yield a trivial dynamics.} shown in Fig.~\ref{fig:SU2_tree}, collapsing the state to either a spin singlet or a spin triplet. Each mid-circuit measurement happens with probability $p$. In all, our model has three tunable parameters: the two angle parameters $\theta_1$ and $\theta_2$, which determine the entangling gates, and the mid-circuit measurement probability $p$. We are interested in whether the two probe qubits exhibit purification and spin-sharpening transitions upon tuning these parameters.

In all cases of $SU(2)$-symmetric dynamics we considered, an initially mixed state has a finite probability of remaining mixed at infinitely late time, meaning there is no pure phase. This is unlike the one-dimensional case, where mixed states asymptotically purify~\cite{majidy_critical_2023}. 
Additionally, for $p < 1$ we find that the system does not sharpen either (i.e. there is a finite probability of spin fluctuations surviving to infinite time). 
For $p = 1$ we do find sharpening in certain ranges of the gate angles $\theta_1, \theta_2$. We are able to find the phase boundaries for this unexpected transition by an explicit analytic argument (which agrees well with finite-size numerics). Furthermore, our theoretical derivation in Appendix.~\ref{append:SU2_scale} suggests that this phase transition also belongs to the glass universality class \cite{nahum_measurement_2021,feng_measurement_2023,Derrida_Polymers_1988}. Our finding of a spin sharpening phase transition subject to fine-tuning of $p=1$ is quite different from the case of one-dimensional models~\cite{majidy_critical_2023}, where a transition from a fuzzy phase to a critical phase is observed without fine-tuning.

Before we end this section, we want to comment on the difference between phase transitions in the tree model and those in one-dimensional circuits~\cite{majidy_critical_2023}. In one dimension, the purification (respectively, sharpening) transition can be characterized by the timescales for the system to be purified (respectively, sharpened). 
The rate of decay of the quantity of interest $S$ (which can be either the entropy or charge fluctuation) takes the form $\dot{S} \simeq -S/\tau(N)$, with $N$ the system size, giving the late-time exponential decay $S(t) \sim e^{-t/\tau(N)}$. Here $\tau$ is a function of system size $N$, that could be constant (sharp phase---not seen in one dimension~\cite{majidy_critical_2023}), algebraic (critical phase) or exponential (fuzzy phase). Either way $S\to0$ at infinitely late time. 
On the contrary, in tree model, $N=2^t$, so the late time decay of $S(t)$ must be a function of $t$ alone.
A na\"ive attempt to solve the decay $\dot{S} = -S/\tau(N)$ with $N = 2^t$ gives that $S$ saturates to a finite nonzero value (when $\tau=\exp(N)$ in the mixed/fuzzy phase) or decays to $0$ exponentially with $\tau = O(1)$ timescale (pure/sharp phase). 
This reasoning suggests a different phenomenology for the tree compared to one-dimensional (or generally Euclidean) models, which we study below. It also leads us to use the value of the order parameter (entropy or charge fluctuation) at infinite time, rather that its decay timescale, to characterize the phase transition in the tree model. 

\section{Quantum tree with $U(1)$ symmetry}
\label{sec:U1}

In this section we study the purification and sharpening transitions in the collapse tree model with $U(1)$ symmetry, described in Sec.~\ref{sec:keys}. 
We locate the critical points for purification and sharpening analytically in the two limits $d = 1$ (with just a qubit on every site) and $d = \infty$ (with an infinitely large neutral qudit attached to each qubit). For $d = 1$, because of the way the tree model is specified, the purification and sharpening transitions automatically coincide \footnote{We explain this in detail in next subsection.}; by contrast, for $d = \infty$, they occur at well-separated values of $p$. It is natural to ask if these transitions are separate at intermediate values of $d$. We provide clear evidence that they are separate even for the lowest nontrivial values $d = 2$ and $3$.

\subsection{Phase transition for $d=1$}
\label{sec:U1_quant}
In this subsection, we focus on the case $d=1$. In this case the gate $U$ takes the form
\begin{equation}
    U =
    \begin{pmatrix}
        e^{i\phi_0}&\\
        &u^1_{2\times 2}&\\
        &&e^{i\phi_2}
    \end{pmatrix},
\end{equation}
with $\phi_{0,2}$ uniformly random in $[0,2\pi)$ and $u^1$ a Haar random $2\times 2$ unitary matrix in the charge sector with total charge $1$.
In the collapse tree, the density matrix is always a $2\times 2$ matrix in each time step. This allows us to use the method in Refs.~\cite{Derrida_Polymers_1988,nahum_measurement_2021,feng_measurement_2023} to study the phase transitions recursively.  

First, we briefly comment on why the transitions must coincide. Consider the purification setup, in which the initial state is maximally mixed. This state contains no coherences between distinct $U(1)$ sectors, and the dynamics (which conserves $U(1)$) cannot generate such coherences. Suppose we are in the pure phase, so the qubit at the top of the tree is in a pure state. The only pure states without coherences between charge sectors are the charge basis states $\ket{0}, \ket{1}$, which are sharp. Therefore, for a maximally mixed initial state, purification and sharpening must coincide. Moreover, for the symmetric random ensemble of gates we have specified, the measurement outcomes are independent of the phase between $\ket{0}$ and $\ket{1}$ on any site, so the sharpening of incoherent mixtures is equivalent to that of coherent superpositions. So we conclude that the sharpening and purification transitions must coincide, and use the ``purification'' initial condition of a maximally mixed state as it permits an exact solution.

With this initial condition there is generally a mixed state for the outgoing qubit at each node. The corresponding density matrix is strictly diagonal in the $z$ basis since we have the $U(1)$-symmetry around the $z$ direction and, as mentioned above, no coherences between charge sectors are created in the dynamics. This reduces the amount of information we need to keep track of during the dynamics. Similar to previous work \cite{Derrida_Polymers_1988,nahum_measurement_2021,feng_measurement_2023}, we define the order parameter $\mathcal{Z}^{typ}$
\begin{equation}
    \ln \mathcal{Z}^{typ} =\langle \ln \mathcal{Z}\rangle_{\ne 0},
\end{equation}
with $\mathcal{Z}$ as the smaller eigenvalue of density matrix, and $\langle \cdots \rangle$ represents the averaging over measurement outcomes and unitary gates. The subscript $\langle\cdots\rangle_{\neq 0}$ indicates that only the nonzero values of $\mathcal{Z}$ are included in the average. At $p=1$, the system is purified immediately so that $\mathcal{Z}=0$ with probability $1$. This pathology in the definition can be fixed by letting $\mathcal{Z}^{typ}=0$ at $p=1$. If $\mathcal{Z}^{typ}$ is nonzero when the number $k$ of layers of the tree goes to infinity, it means that there is a finite probability of having a mixed final state. This represents the mixed phase. Otherwise, the final state becomes pure with probability $1$ as $k\to\infty$. That means we are either in the pure phase or at the critical point. 
In addition to the $\mathcal{Z}$ value, another bit of information needed to fully label the density matrix is which eigenvector has smaller eigenvalue. Since the $U(1)$ symmetry guarantees that the density matrix is diagonal in the $z$ basis, this is equivalent to knowing whether the smallest eigenvalue is associated to the eigenvector $\ket{0}$ or $\ket{1}$. We use an extra bit $s=0,1$ to represent that the density matrix is closer to $\ket{s}$. So generally for each subtree, we label the density matrix of the top qubit by the pair $(\mathcal{Z},s)$, which corresponds to density matrix 
\begin{equation}
    \rho = (1-\mathcal{Z})\ket{s}\bra{s}+\mathcal{Z}\ket{1-s}\bra{1-s}\label{eq:U(1)_densitym}.
\end{equation}
Thus the ensemble of quantum trajectories (across both measurement outcomes and gate realizations) of a depth-$k$ tree is fully specified by a probability distribution $p_k(\mathcal{Z},s)$ over $[0,1/2]\times \{0,1\}$. However, we can further notice that the initial (maximally mixed) state has the $\mathbb{Z}_2$ symmetry of flipping all qubits, and that the circuit ensemble is also statistically invariant under this symmetry (i.e., any given realization of the dynamics is mapped to another equally likely realization). This symmetry gives the condition $p_k(\mathcal{Z},s=0)=p_k(\mathcal{Z},s=1)$. It implies that we only need to record the distribution $p(\mathcal{Z})=p(\mathcal{Z},s=0)+p(\mathcal{Z},s=1)$ during the dynamics\footnote{In the numerical calculation, we still keep track of the pair $(\mathcal{Z},s)$, but in the analytical discussion we utilize the property $p(\mathcal{Z},0)=p(\mathcal{Z},1)$.}. 

\begin{figure*}
    \centering
    \includegraphics[width=1.0\textwidth]{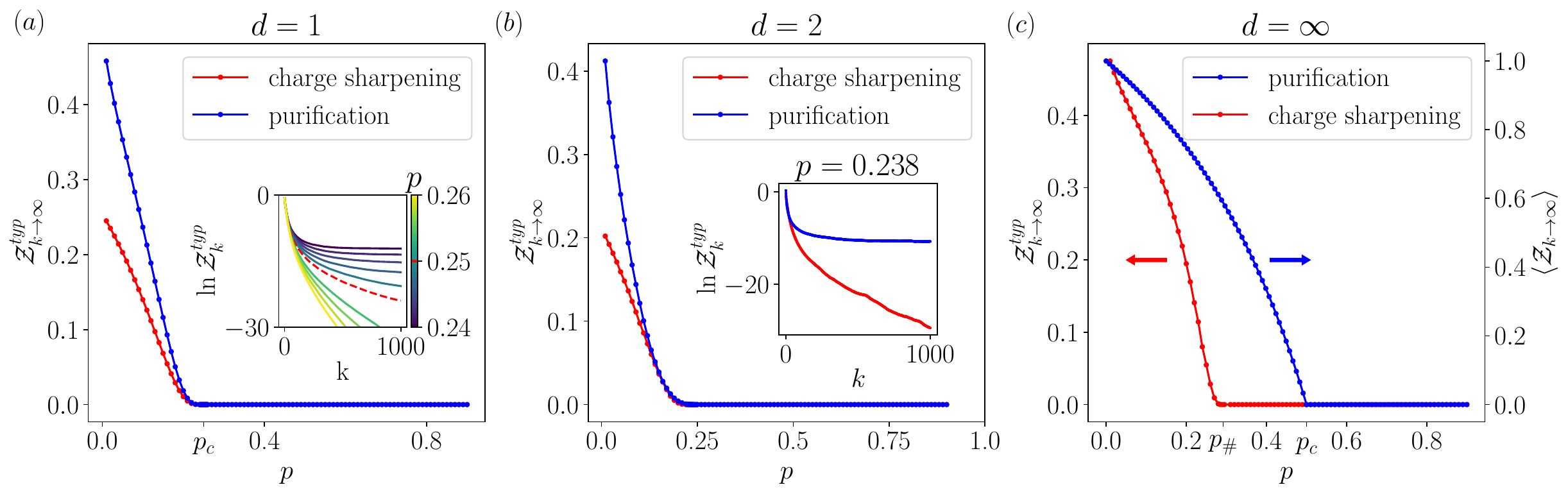}
    \caption{
    Numerical simulations of the collapse tree with $U(1)$ symmetry.
    (a) $d=1$. The purification and charge sharpening transitions are identical in this case. Main panel: infinite-depth limit of $\mathcal{Z}^{\rm typ}$ starting from a fully-mixed state (`purification') and a pure $X$-basis product state (`charge sharpening'). Both cases show a transition at $p_c=0.25$. 
    Inset: dynamics of $\ln Z^{typ}_k$ in the vicinity of the charge sharpening transition, $0.24 \leq p \leq 0.26$. The curve with $p=0.25$ is shown by the red dashed line. All results are obtained by the pool method with pool size $10^7$. 
    (b) Same plot for $d=2$. The pool size is $10^6$. Inset: the dynamics of $\ln Z^{typ}_k$ at fixed $p=0.238$ indicates that the system is in a mixed, charge-sharp phase, showing a separation between the two critical points. 
    (c) $d\to\infty$ limit. The critical point of purification transition $p_c=0.5$ and the critical point of charge sharpening transition $p_{\#}=1-\frac{\sqrt{2}}{2}$ are indicated on the $x$ axis. The pool size is $10^7$. The purification curve is obtained by analytical solution of the percolation transition of binary tree, as discussed in Appendix~\ref{append:lin_recur_U1}. In the purification transition with $d=\infty$, the order parameter $\langle\mathcal{Z}_{k\to\infty}\rangle$ is defined as the average length of domain wall. Across all panels, error bars are smaller than the size of the markers.}
    \label{fig:U1_transitions}
\end{figure*}

A powerful approach to study the dynamics of our quantum tree model is based on its recursive structure. In Appendix~\ref{append:recur_struct_U1}, we give a detailed discussion to show that the dynamics of quantum trees obeying $U(1)$-symmetry can be simulated recursively. This enables the so called \textit{pool method} \cite{Miller_Weak-disorder_1994,Monthus_Anderson_2009,Garcia-Mata_Scaling_2017} which simulates the distribution of density matrices at layer $k$ from a pool of density matrices at layer $k-1$, allowing the efficient simulation of trees of large size. Numerical results for the phase transitions at $d=1$ are shown in Fig.~\ref{fig:U1_transitions}(a). Note that no approximation is used. The pool size is large enough to guarantee that the curve converges. From the numerical results, we see a phase transition from the mixed (fuzzy) phase to the pure phase (sharp) when $p$ is increased. 

In order to understand the phase transition analytically, we analyze $\mathcal{Z}$ recursively. The key idea is that for a node at layer $k+1$, we can write down the recursive relation between the $\mathcal{Z}$ value of the $(k+1)$-layer tree and the $\mathcal{Z}_1$, $\mathcal{Z}_2$ values of the two $k$-layer subtrees that meet at step $k+1$. The recursive relation is controlled by the unitary gate and measurement outcomes in the node where the subtrees meet. In addition to the recursive relation of $\mathcal{Z}$ for each possible node configuration, we also need the probabilities of different measurement outcomes inside the node \cite{feng_measurement_2023}, which can be obtained from the Born rule. To solve the dynamics, we further introduce the generation function $G_k(x)= \langle\exp(-e^{-x}\mathcal{Z}_k)\rangle$. It can be naively viewed as a smeared version of a step function with argument $x$. When $x\gg \ln \mathcal{Z}_k^{typ}$, $G_k(x)$ is a plateau at $1$ and when $x\ll \ln \mathcal{Z}_k^{typ}$, it is a plateau at zero. $G_k(x)$ only changes dramatically near $x = \ln \mathcal{Z}_k^{typ}$. It is convenient to view $k$ as discrete time and $x$ as position\footnote{Note this is a fictitious space, not related to the physical system.}. Then $G_k(x)$ describes a moving wave with the front located at $\ln \mathcal{Z}_k^{typ}$. In the mixed phase, where $\mathcal{Z}_k^{\rm typ}$ remains finite at $k\to\infty$, the wave described by $G_k(x)$ stops moving after a sufficiently long time. However, in the pure phase, the wave front keeps moving toward $x=-\infty$ with a negative velocity $v$. At the critical point, the velocity vanishes, $v=0$. Since after long enough time, $\mathcal{Z}_k^{typ}$ becomes arbitrarily small in the pure (sharp) phase and at the critical point, the effect of higher-order terms in $\mathcal{Z}$ is suppressed, so it is sufficient to keep only the leading-order terms of the recursive relation to get the velocity. 

We leave the derivation of the velocity in Appendix~\ref{append:lin_recur_U1} and here we only give the final results. In short, plugging the linearized recursive relations in the definition of $G_k(x)$ gives a Fisher-KPP-like recursive relation. This has been well studied in Ref.~\cite{Derrida_Polymers_1988}. By leveraging those results, we find that the purification transition happens at  $p_c=1/4$. Compared to numerical result in Fig.~\ref{fig:U1_transitions}(a), we see that the analytical result matches well with our numerical calculation, confirming the validity of our theory. 
 
In addition to the value of $p_c$, it is also interesting to obtain the scaling exponents characterizing the dynamics near the critical point. In Appendix~\ref{append:lin_recur_U1}, we point out that the universality class is still the so called glass class in previous work \cite{nahum_measurement_2021,feng_measurement_2023}, where the scaling behaviors are
 \begin{align}
     \mathcal{Z}^{typ}_{k\to+\infty}& \sim \exp(-\frac{C}{\sqrt{p_c-p}}) & (p & \lesssim p_c ), \nonumber\\
     \ln \mathcal{Z}^{typ}_k& \sim -k^{1/3} & (p & =p_c).
     \label{eq:U1_scale}
 \end{align}
We have also studied the dynamics from an initial state that is a superposition rather than a mixture (see Appendix~\ref{append:lin_recur_U1} for more details). In this case we only have a numerical solution, but this solution gives an estimate of the critical point that is consistent with our analytical prediction above, see Fig.~\ref{fig:U1_transitions}(a). 

\subsection{Phase transitions in large-$d$ limit}
\label{sec:U1_class}
In the limit $d\to\infty$, our model can be mapped to a classical statistical-mechanical model where quantum frustration is suppressed. The entanglement entropy is determined by the minimal length of a domain wall separating the initial and final boundary condition \cite{Zhou_Emergent_2019,Skinner_2019,Bao_Theory_2020}. Then the purification transition becomes a classical percolation problem. When the dynamics has $U(1)$ symmetry, previous work \cite{agrawal_entanglement_2022} shows that the $n$-th R\'enyi entropy takes the form
\begin{equation}
    S_n (\mathbf{m}) = S_n^T(\mathbf{m})+l_{DW}\ln d, \label{eq:renyi_ansatz}
\end{equation}
where $\mathbf{m}$ represents the location of all measurements and the outcome of measurements on the qubits\footnote{Notice that $\mathbf{m}$ contains only the measurement outcome of the qubit degree of freedom, not of the neutral qudits.}, while $l_{DW}$ is the minimal length of the domain wall. The extra term $S_n^T$ can be understood as the charge configuration entropy on the domain wall, which characterizes the charge sharpening transition in the limit $d\to \infty$ for a fixed choice of $\mathbf{m}$. The average R\'enyi entropy is obtained by averaging over all possibilities of $\mathbf{m}$. The form Eq.~\eqref{eq:renyi_ansatz} of the R\'enyi entropies greatly simplifies the study of the charge sharpening and purification transitions. In the rest of this subsection, we give both analytical solution and numerical results of these two phase transitions in our tree model.

The purification transition is determined by the minimal domain wall length $l_{DW}$. In our problem, this is just percolation on a tree which gives $p_c= 1/2$ (see Appendix~\ref{append:lin_recur_U1}). Considering that the minimal domain wall has length $\leq 1$ (it is always possible to insert a length-1 domain wall immediately below the top qubit), we define $\mathcal{Z}=0$ when the tree is disconnected ($l_{DW} = 0$) and $\mathcal{Z}=1$ when the tree is connected ($l_{DW}=1$). This allows us to use $\langle \mathcal{Z}_{k\to\infty}\rangle$, which is the average length of the minimal domain wall in the $k\to\infty$ limit, as the order parameter. Results are shown in Fig.~\ref{fig:U1_transitions}(c). As for the extra term $S_n^T$ in Eq.~\eqref{eq:renyi_ansatz}, it takes the form~\cite{agrawal_entanglement_2022}
\begin{equation}
    S_n^T(\mathbf{m}) = -\frac{1}{n-1}\ln \left(\sum_{\{\beta\}} p_{\{\beta\}}^n\right),
\end{equation}
with $p_{\{\beta\}}$ the probability that the charge configuration on the unmeasured links of the domain wall is $\{\beta\}$. For the tree problem, as mentioned previously, we have $l_{DW}\leq 1$: the top and bottom of the tree can always be separated by at most one cut (e.g. right below the top qubit). Thus, in the pure phase, the tree is disconnected ($l_{DW}=0$) and $S_n^T$ should be zero, since there are no charge configurations to sum over. In the mixed phase, the tree is generally connected ($l_{DW}=1$); simply choosing to cut the link on the top of the tree, we have that $S^T_n$ measures the uncertainty on the charge in the final state. This can be solved analytically by considering the recursive structure of the tree. We leave the details in Appendix~\ref{append:lin_recur_U1} and only summarize the main ideas in this subsection.

Each subtree can be represented by a vector which measures the probabilities of the top qubit (of the subtree) to be in state $\ket{0}$ or $\ket{1}$. This allows us to introduce a pair $(\mathcal{Z},s)$ to represent the vector\footnote{As discussed in Ref.~\cite{feng_measurement_2023}, using $\mathcal{Z}$ is equivalent to using the Renyi entropy $S^{T}_n$ with arbitrary $n$ as the order parameter.}. Here $\mathcal{Z}\leq 1/2$ is the smaller of the two probabilities and $s=0,1$ denotes the state with larger probability.  
The initial state is a uniform distribution over all charge configurations. In this case, too, the tree structure enables a recursive analysis. This charge sharpening transition is determined by studying the recursive relation of $(\mathcal{Z},s)$, similar to the $d=1$ case. One thing to note in relation with the $d=1$ case is that, even though the initial state in this case is a coherent superposition rather than a mixed state, the dynamics can still be studied in terms of classical probability distributions. Namely each node outputs a distribution given two distributions from the two incoming subtrees as input. We can still define $\mathcal{Z}^{typ}$ as 
\begin{equation}
     \ln \mathcal{Z}^{typ} =\langle \ln \mathcal{Z}\rangle_{\ne 0}.
\end{equation}
Within each node, we have two input pairs $(\mathcal{Z}_1,s_1)$ and $(\mathcal{Z}_2,s_2)$ to characterize the charge of the input subtrees.
In the case $d=1$, we have a random entangling gate acting on the two input states. 
Here instead, the entangling gate within each node is replaced by a transition matrix acting on the input distribution \cite{agrawal_entanglement_2022}
\begin{equation}
    V = 
    \begin{pmatrix}
        1&0&0&0\\
        0&1/2&1/2&0\\
        0&1/2&1/2&0\\
        0&0&0&1
    \end{pmatrix}.
\end{equation}
After this transition matrix, we measure one site and obtain the pair $(\mathcal{Z},s)$ that characterizes the distribution of charge on the remaining site. From these steps, we can write the recursive relation for $(\mathcal{Z},s)$.
The method of Ref.~\cite{Derrida_Polymers_1988} can still be used to study the dynamics of $\mathcal{Z}^{typ}$. For all leaves of the tree, the initial probability vector is $(1/2,1/2)^T$. This guarantees the $\mathbb{Z}_2$ charge-conjugation symmetry as before. By studing the dynamics of the generating function $G_k(x)=\langle \exp(-e^{-x}\mathcal{Z}_k)\rangle$, we find that the charge sharpening transition happens at $p_{\#}=1-\frac{\sqrt{2}}{2} \simeq 0.293$, which deviates from the classical percolation transition $p_c=1/2$. We further find that the scaling exponents of the charge sharpening phase transition are also described by Eq.~\eqref{eq:U1_scale} (see details in Appendix.~\ref{append:lin_recur_U1}). 

The charge sharpening transition can be simulated efficiently by the pool method. Our results are shown in Fig.~\ref{fig:U1_transitions}(c). In the fuzzy phase, $\mathcal{Z}^{typ}_{k\to\infty}\ne 0$ while $\mathcal{Z}^{typ}_{k\to 0}=0$ in the sharp phase. We see that the charge sharpening transition happens at a value consistent with the analytically computed $p_{\#}=1-\frac{\sqrt{2}}{2}$, and clearly far from the purification transition critical point $p_c = 1/2$.

\subsection{Critical points separation}
\label{sec:critical_points_separation}

In the previous discussion, we showed that the critical points of the charge sharpening and purification transitions coincide for $d = 1$, while in the limit $d \to \infty$ the critical points flow to $p_c=1/2$ and $p_{\#}=1-\frac{\sqrt{2}}{2}$ respectively. This suggests that these two critical points separate with the increase of $d$. Then it is interesting to check what happens for finite $d>1$. We show numerical results for the purification and charge sharpening transitions for $d=2$ in Fig.~\ref{fig:U1_transitions}(b). For the purification transition, we define $\mathcal{Z} = 1-\mu$, with $\mu$ the largest eigenvalue of the final density matrix~\footnote{In the case $d=1$, this reduces to our definition of $\mathcal{Z}$ as the smaller eigenvalue.}. For the charge sharpening transition, we set $\mathcal{Z} = {\rm min}(p_0,p_1)$ with $p_s$ the total probability of charge sector $s=0,1$ in the final state. The reason why we no longer use $\mathcal{Z}$ as the order parameter is because we can generally have determined charge while the $d$-level ancilla qubit remains mixed. In the inset, we give the dynamics of $\ln \mathcal{Z}^{typ}_k$ at $p=0.238$. Although the curve for the purification process saturates to a finite value, the curve for the charge sharpening process decays toward $-\infty$. This suggests that $p_{\#} < 0.238 < p_c$, confirming the idea that the critical points separate with the increase of $d$. It is worth noting that our results also suggest that the critical points move to lower $p$ when going from $d=1$ to $d=2$. Combined with our previous results on $d\to\infty$ (where both $p_\#$ and $p_c$ are above 0.25), this observation implies an unexpected non-monotonic dependence of the critical points on $d$. 

To better understand the behavior of critical points $p_c$, $p_\#$ as a function of $d$, we further check the purification and charge sharpening transitions with $d=3$ numerically (see Appendix~\ref{appendix:U1_d3} for more details). We summarize the location of critical points $p_c$, $p_\#$ for $d=1,2,3,\infty$ in Table~\ref{tab:critical_points}. For the transitions that lack an analytical understanding, we give lower and upper bounds based on our numerical simulations. We can see from Table~\ref{tab:critical_points} that the critical points move to larger $p$ at $d=3$, supporting our expectation that they should eventually flow to $p_c=1/2$ and $p_{\#}=1-\frac{\sqrt{2}}{2}$ with the increase of $d$. An analytical understanding of the counterintuitive decrease from $d=1$ to $d=2$ remains as an open problem for future work.

\begin{table}
    \centering
    
    \begin{tabular}{ccccc}
    \toprule
         &$d=1$&$d=2$&$d=3$&$d=\infty$  \\
         \midrule
        purification & 
        $0.25$ & 
        $0.243(3)$ &  
        $0.305(5)$ & 
        $0.5$\\ 
       
        charge sharpening & 
        $0.25$ &
        $0.229(5)$ & 
        $0.243(4)$ & 
        $1-\frac{\sqrt{2}}{2}\simeq 0.293$ \\
        \bottomrule
    \end{tabular}
    \caption{Critical points of purification and charge sharpening transitions for different values of the Hilbert space dimension $d$ of the neutral degree of freedom. Results with $d=1,\infty$ are exact, while for $d=2,3$ they are obtained from numerical simulations, with the uncertainties shown.}
    \label{tab:critical_points}
\end{table}

\section{Quantum tree with $SU(2)$ symmetry}
\label{sec:spin_sharpening}

%\subsection{Model}
In the previous section, we discussed the dynamics of a collapse quantum tree with $U(1)$ symmetry. Here we turn to the dynamics of a tree model with $SU(2)$ symmetry. Previous work shows that the phenomenology of monitored dynamics in one-dimensional systems with $SU(2)$ symmetry is quite rich and different from that of the $U(1)$ case \cite{majidy_critical_2023}. However, the non-Abelian nature of the $SU(2)$ symmetry makes it hard to get an analytical understanding of the dynamics and the associated phase diagram. In this section, we consider the expansion tree model metioned in Sec.~\ref{sec:keys} with $SU(2)$ symmetry in which the purification and spin-sharpening transitions are more tractable, allowing us to obtain the first analytical results on spin sharpening in monitored many-body dynamics.

At a glance, it seems natural to first study the purification transition. However, the $SU(2)$ symmetry places a strong constraint on purification. Namely, one of the possible outcomes of spin sharpening, the spin triplet, is still a mixed state: $\rho = P_t / 3$. Therefore, a pure phase requires that (almost) all trajectories sharpen to a spin singlet in the $k\to\infty$ limit. This clearly violates spin conservation\footnote{The fully-mixed input state has $\langle P_t\rangle = 3/4>0$, while the monitored trajectories in this scenario would have on average $\langle P_t \rangle = 0$.}, and is therefore not allowed in the case of `real' measurements, as we show in more detail below. Thus, the system always stays in the mixed phase. For this reason, in the rest of this section we mainly focus on the problem of spin sharpening.

\begin{figure}
    \centering
    \includegraphics[width=1.0\columnwidth]{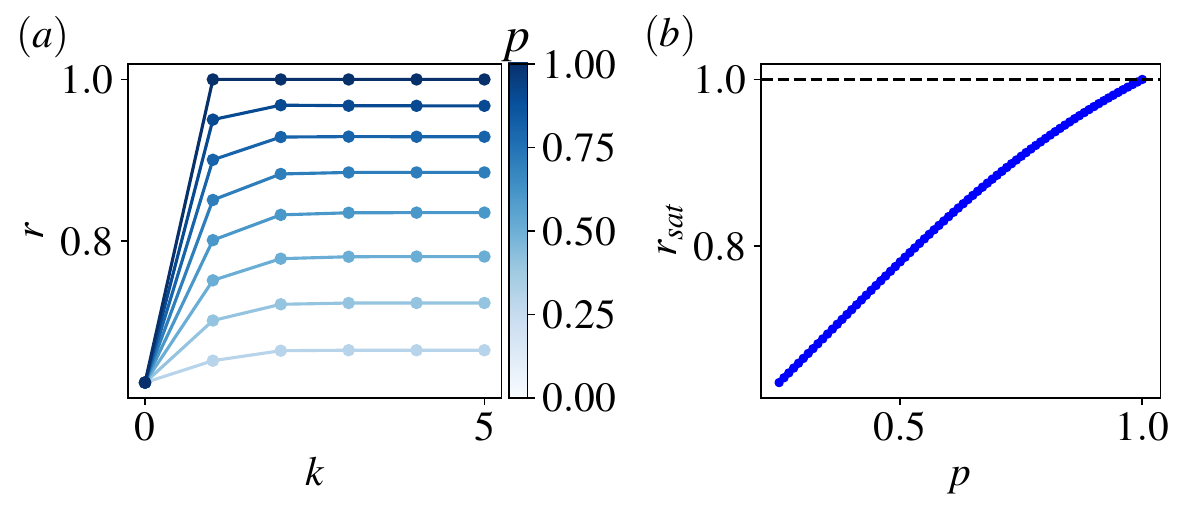}
    \caption{Spin-sharpening dynamics on $SU(2)$-symmetric tree for gate angles $\theta_1=\theta_2=\pi/2$. We use the order parameter $r$ defined in Eq.~\eqref{eq:SU2_order} to characterize the phases (the system is sharp if and only if $r=1$).
    (a) The dynamics of $r$ vs tree depth $k$, for different measurement probability $p$. 
    (b) Saturation value value of $r$ (at large $k$) vs measurement probability $p$.}
    \label{fig:SU2_fixgate}
\end{figure}

One important property of the dynamics with $SU(2)$ symmetry is that the reduced density matrix of the two reference qubits takes the form
\begin{equation}
    \rho_{\mathbf m} = \sigma(\mathbf m) P_s+\frac{\tau(\mathbf m)}{3}P_t,
\end{equation}
with $\sigma(\mathbf{m})$ and $\tau(\mathbf{m})$ some non-negative real numbers. In this section we use the un-normalized density matrix obeying ${\rm Tr}(\rho_{\mathbf{m}}) = \sigma(\mathbf{m}) + \tau(\mathbf {m})\equiv p( \mathbf{m})$, the probability of getting the specific trajectory $\mathbf{m}$. 
With this convention, the coefficients $\sigma(\mathbf{m})$ and $\tau(\mathbf {m})$ have clear physical interpretations: they are the joint probabilities of obtaining mid-circuit outcomes $\mathbf{m}$ {\it and} measuring the reference qubits in a singlet or triplet state, respectively.
It follows that the conditional probability of finding the references in a singlet state given trajectory $\mathbf{m}$ is $\sigma(\mathbf{m}) / [\sigma(\mathbf m) + \tau(\mathbf m)]$, and similarly for the triplet. Spin sharpening can then be characterized, for example, in terms of the order parameter $(\sigma^2+\tau^2)/(\sigma+\tau)^2$, which equals 1 if and only if the trajectory is spin-sharp (i.e., either $\sigma = 0$ or $\tau = 0$), and is below 1 otherwise. 
Then we define the quantity
\begin{align}
    r & = \sum_{\mathbf{m}} \left( \frac{\sigma(\mathbf{m})+\tau(\mathbf{m})}{\sum_{\mathbf{m}'}\sigma(\mathbf{m}')+\tau(\mathbf{m}')} \right)\frac{\sigma(\mathbf{m})^2+\tau(\mathbf{m})^2}{(\sigma(\mathbf{m})+\tau(\mathbf{m}))^2} 
    \label{eq:SU2_order}
\end{align}
to serve, in the $k\to\infty$ limit, as an order parameter\footnote{We can still define the order parameter $\mathcal{Z}^{typ}$ like in the $U(1)$ case, which is used in the theoretical approach in Appendix~\ref{append:SU2_lin_recur}. But $r$ can be easily generalized to the case of non-Born-rule measurement dynamics in Appendix~\ref{append:fake_measurement}} for spin sharpening on average across trajectories. Since the gates are fixed, the summation over $\mathbf{m}$ contains all realizations of the tree. 

Before moving on, we want to give an illustration of the dynamics of $r$. To do so, we consider a special case with $\theta_1=\theta_2=\pi/2$, which represents SWAP gates. Then it is easy to see from Fig.~\ref{fig:SU2_tree} that, when $p=1$, the system becomes spin-sharp immediately after the first time step. 
It turns out that, even when $p<1$, the evolution of $r$ saturates very quickly to a finite value $r<1$, indicating a fuzzy phase. 
In Fig.~\ref{fig:SU2_fixgate}, we show the behavior of $r$ vs $k$ for different values of $p$. It can be seen that the sharp phase is present only at $p=1$, and the late-time value of $r$ drops below 1 immediately as $p<1$. This simple example already illustrates a general aspect of $SU(2)$-symmetric monitored dynamics, which is the prevalence of the fuzzy phase and the difficulty in quickly sharpening the spin. This is also reminiscent of previous results for 1D systems~\cite{majidy_critical_2023}. In the following, we move away from the simple $\theta_1 = \theta_2 = \pi/2$ point and develop more general methods to explore the broader phase diagram in $(p,\theta_1,\theta_2)$.

\subsection{Recursive construction of trajectory ensemble}
The main challenge in our problem is the exponential growth of system size with the tree depth $k$. This requires us to consider the wavefunction of very large systems\footnote{Although our target is only the density matrix of probe qubits, we need to keep track of the wavefunction of all qubits during the dynamics.} even for modest depths $k$. To overcome this obstacle, we aim to use the structure of the tree to get the a recursion relation for the density matrices, like we have done in the $U(1)$ case in Sec.~\ref{sec:U1}. We leave all details in Appendix~\ref{append:SU2_recur} and just give a summary of the main results here. 

We can view an arbitrary expansion tree with $k$ layers as connecting two expansion subtrees with $k-1$ layers, via a single node at the bottom (see Fig.~\ref{fig:SU2_tree}). Suppose we know the reference density matrices of the two subtrees; we can then recursively obtain the reference density matrix of the whole expansion tree. 
Labeling the two subtrees connecting to the bottom node by $(\sigma',\tau')$ and $(\sigma^{\prime\prime},\tau^{\prime\prime})$, we can write down the values of $(\sigma,\tau)$ for the whole tree as follows:
\begin{align}
    \sigma =& c^s_{ss}(\theta_1,\theta_2,\tilde{\mathbf{m}})\sigma'\sigma^{\prime\prime}+c^s_{tt}(\theta_1,\theta_2,\tilde{\mathbf{m}})\tau'\tau^{\prime\prime}\nonumber\\
    \tau=& c^t_{st}(\theta_1,\theta_2,\tilde{\mathbf{m}})(\sigma'\tau^{\prime\prime}+\tau'\sigma^{\prime\prime})+c^t_{tt}(\theta_1,\theta_2,\tilde{\mathbf{m}})\tau'\tau^{\prime\prime}\label{eq:sigma_tau_recur}.
\end{align}
These are the bilinear recursive relations for the parameters $\sigma$ and $\tau$. The coefficients $c$ depend on the details of the node, including the gate angle parameters and outcomes $\tilde{\mathbf{m}}$ of the two measurements inside the node. It can be noticed that Eq.~\eqref{eq:sigma_tau_recur} does not contain all possible quadratic combinations of $\sigma$ and $\tau$: missing terms are not allowed by the symmetry\footnote{This can be seen using the  Clebsch–Gordan coefficients.}. 

Eq.~\eqref{eq:sigma_tau_recur} in principle also yields a recursive relation for the probability distribution over measurement outcomes, $p = \sigma + \tau$. However, this can not be simply simulated using the pool method\footnote{In fact, our recursive relation Eq.~\eqref{eq:sigma_tau_recur} is equivalent to viewing the expansion tree as a time-reversed collapse tree; however, the singlet pairs injected into each node map under time reversal to \textit{post-selected} spin measurements whose outcome is forced to be a singlet. Then the corresponding collapse tree contains a mix of real and forced measurements. As we show in Appendix~\ref{append:SU2_recur}, the distribution of outcomes cannot be simulated recursively in this case.}. This forbids us to use the pool method to simulate large enogh system size. We leave the numerical results to be discussed in next subsection. Here we give some analytical understanding about the dynamics using the recursion relation we have. 

First, looking at Eq.~\eqref{eq:sigma_tau_recur}, it can be seen that when $c^s_{tt}$ and $c^t_{tt}$ are both nonzero, the output density matrix is generally a fuzzy state. This is because, even when both subtrees have fully sharpened into triplets ($\sigma' = \sigma'' = 0$), their merger is a mixture of singlet and triplet ($\sigma,\tau>0$). 
The only possibility to get a sharp phase would be a case with singlet states alone. However, this is not possible. If we sum the quantity $\tr(\rho_{\mathbf{m}}P_s)$ over all trajectories, we get
\begin{equation}
    \sum_{\mathbf{m}}p(\mathbf{m})\tr(\rho_{\mathbf{m}}P_s)=\tr\left(\sum_{\mathbf{m}}p(\mathbf{m})\rho_{\mathbf{m}}P_s\right)=\frac{1}{4}\label{eq:prob_constrain}
\end{equation}
(the last equality uses the fact that when we sum the density matrix over all trajectories based on their probability, we get the maximally mixed state). A sharp phase with (almost) all states $\rho_{\mathbf m} \propto P_s$ would give $\tr(\rho P_s) = 1$, violating Eq.~\eqref{eq:prob_constrain}. In fact, the constraint Eq.~\eqref{eq:prob_constrain} requires us that if there is a sharp phase, then $1/4$ of the trajectories have to sharpen to the spin singlet and $3/4$ of trajectories have to sharpen to the spin triplet. 
Our calculation of the $c$ coefficients shows that when $p<1$, we generally have $c^s_{tt}$ and $c^t_{tt}$ both nonzero simultaneously (see Appendix~\ref{append:SU2_recur} for more details). It follows that there is no sharp phase when $p<1$, generalizing our finding for the special point $\theta_1 = \theta_2 = \pi/2$ (Fig.~\ref{fig:SU2_fixgate}) to the whole $\theta_{1,2}$ plane. Moreover, we note that this property still holds true even in more general gate sets obeying $SU(2)$ symmetry. For instance, when we have randomness in $\theta_1$ and $\theta_2$ across different nodes of the tree, we can still use the same argument to show there can be no sharp phase except at $p=1$.

To get a sharp phase, we turn to the dynamics at $p=1$ where we find that at least one of $c^s_{tt}$ and $c_{tt}^t$ becomes zero (see Table~\ref{tab:output_SU2} in Appendix.~\ref{append:SU2_recur}). Setting $p=1$, we are interested in understanding whether a sharp phase is allowed by tuning parameters $\theta_1$ and $\theta_2$.
In Appendix~\ref{append:SU2_lin_recur}, we present a theoretical approach to this problem. Similar to the $U(1)$ case,  
we introduce the concept of $\mathcal{Z}$ value set to be the smaller value between $\sigma/(\sigma+\tau)$ and $\tau/(\sigma+\tau)$ (the conditional probabilities of finding the reference qubits in a singlet or triplet state respectively). It can be related to $r$ by $r\approx1-2\langle \mathcal{Z}\rangle$ when $\langle \mathcal{Z}\rangle$ is close to $0$. We also use an extra bit $\eta$ to represent which one between $P_s$ and $P_t$ the density matrix is closer to. Then $\eta$ generally classifies the density matrices into two sets: singlet-like and triplet-like. We introduce two generating functions $G^s_k(x)=\langle \exp(-e^{-x}\mathcal{Z}_k)\rangle_{s}$ and $G^t_k(x)=\langle \exp(-e^{-x}\mathcal{Z}_k)\rangle_{t}$ which average over only the singlet-like and triplet-like trajectories respectively. These two generating functions behave like two moving waves, with fronts located at $\ln\mathcal{Z}_k^{s,typ}=\langle \ln \mathcal{Z}_k\rangle_{s,\ne 0}$ and $\ln \mathcal{Z}_k^{t,typ}=\langle \ln \mathcal{Z}_k\rangle_{t,\ne 0}$ respectively. Stability of a sharp phase requires that $\ln\mathcal{Z}^{s,typ}_k$ and $\ln\mathcal{Z}^{t,typ}_k$ both decay to $-\infty$ with the increase of $k$. Our derivation shows that these two moving waves are coupled with each other, thus they have the same negative velocity $v$ in the sharp phase. 

We find that at $p=1$, there can be both a sharp phase and a fuzzy phase depending on $\theta_1$ and $\theta_2$. These two phases are separated by a
phase boundary in the parameter space of $(\theta_1,\theta_2)$. This boundary is determined by the condition $v=0$, which turns out to be satisfied when
\begin{equation} 
\frac{3 + \sqrt{3}}{4} \left|\sin(\theta_1+\theta_2)\right|\left|\sin(\theta_1-\theta_2)\right| = 1.\label{eq:SU2_critical_contour}
\end{equation}
Besides the critical point, it is also an important task to solve the universality class of the phase transition. In Appendix~\ref{append:SU2_scale}, we give a detailed derivation of the universality class. Here we just summarize the main idea. Previous works~\cite{nahum_measurement_2021,feng_measurement_2023} show that the universality class is determined by considering higher order corrections of recursive relation of $\mathcal{Z}$. 
We first consider the scaling behavior of the saturated value $\mathcal{Z}^{s(t),typ}_{k\to\infty}$ when $(\theta_1,\theta_2)$ reaches critical boundary from the fuzzy side. It turns out that
\begin{align}
    \mathcal{Z}_{k\to\infty}^{s(t),typ} \sim \exp(-\frac{K}{\sqrt{\delta_{\theta}}}).
\end{align}
Here $\delta_{\theta}$ is the minimal distance from the point $(\theta_1,\theta_2)$ to the critical boundary. We give the coefficient $K$ in Appendix.~\ref{append:SU2_scale}. Second, we study how $\mathcal{Z}^{s(t),typ}_{k}$ decays to $0$ on the critical boundary. We have
\begin{equation}
    \ln \mathcal{Z}^{s(t),typ}\sim -k^{1/3},
\end{equation}
on the critical boundary.
These results suggest that the spin-sharpening transition obtained by tuning $\theta_1$ and $\theta_2$ at $p=1$ is still in the glass universality class \cite{nahum_measurement_2021}.

\subsection{Numerical results}

We now turn to numerical simulations to verify the validity of our theoretical predictions on the spin-sharpening transition at $p=1$.
As mentioned before, the pool method fails in this case. This limits us to brute-force simulation of the dynamics, which severely limits the accessible system sizes.
For the tree model, the total system size grows exponentially with the tree depth $k$, which limits us to small $k$. Within accessible numerical resources, we manage to simulate trees with up to $k=6$ layers (i.e. a total of $128$ system qubits and 2 reference qubits at the final time).  This is made possible by using Eq.~\eqref{eq:sigma_tau_recur} (which is an exact construction of the full trajectory ensemble) rather than full simulation of a 130-qubit density matrix, which would be intractable. The collection of $\{(\sigma,\tau)\}$ parameters representing the trajectory ensemble still grows doubly-exponentially in $k$, but is found to grow more slowly than $4^{2^k}$ when grouping equivalent trajectories, which pushes the limit on the accessible $k$.

Since we have established that there is no sharp phase for $p < 1$, we focus on the case $p=1$ and check how the order parameter $r$ depends on the gates parameters $\theta_{1,2}$. 
First, we note a symmetry of the phase diagram: when changing $\theta\to \theta+\pi$, the entangling gate $U$ becomes $-U$. This shows that the phase diagram of $(\theta_1,\theta_2)$ is invariant under shifts along the two axes by $\pi$. So we can limit our parameter space to $\theta_{1,2}\in[0,\pi]$ without loss of generality. In fact we also have other symmetries including the reflections $(\theta_1,\theta_2)\to(\theta_2,\theta_1)$ and $(\theta_1,\theta_2) \to (\pi-\theta_1,\pi-\theta_2)$; see Appendix~\ref{append:SU2_recur} for more details.

\begin{figure}
    \centering
    \includegraphics[width=0.9\columnwidth]{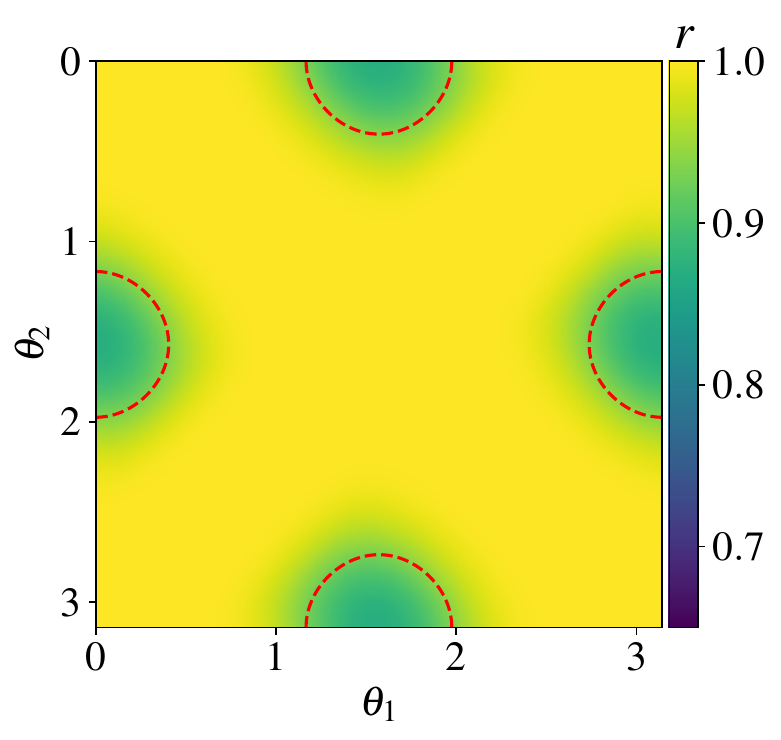}
    \caption{Exact simulation of $SU(2)$-symmetric tree with depth $k=6$, corresponding to a final system of 128 qubits. The spin sharpening order parameter $r$ is shown as a function of gate angles $(\theta_1,\theta_2)$. Our theoretical prediction of the critical boundary with $v=0$, Eq.~\eqref{eq:SU2_velocity}, is shown by the red dashed line.}
    \label{fig:real_128}
\end{figure}

Results of the exact calculation of the $r$ order parameter for $k=6$ (128 qubit final system) are shown in Fig.~\ref{fig:real_128}. The color map represents how sharp the system is on average across trajectories. We see that most of choices of $\theta_1$ and $\theta_2$ give $r$ close to $1$. However, there are also some regions in which the system is still fuzzy in this finite size simulation. In principle this may indicate either a fuzzy phase, or a finite-$k$ transient effect within a sharp phase.
By comparing our theoretical result of the critical boundary Eq.~\eqref{eq:SU2_critical_contour} (red dashed curve) with our computation of the order parameter $r$ in Fig.~\ref{fig:real_128}, we see that our analytical prediction for a fuzzy phase matches well with the observed ``fuzzy'' regions in the numerical simulation.

\begin{figure}
    \centering
    \includegraphics[width=0.9\columnwidth]{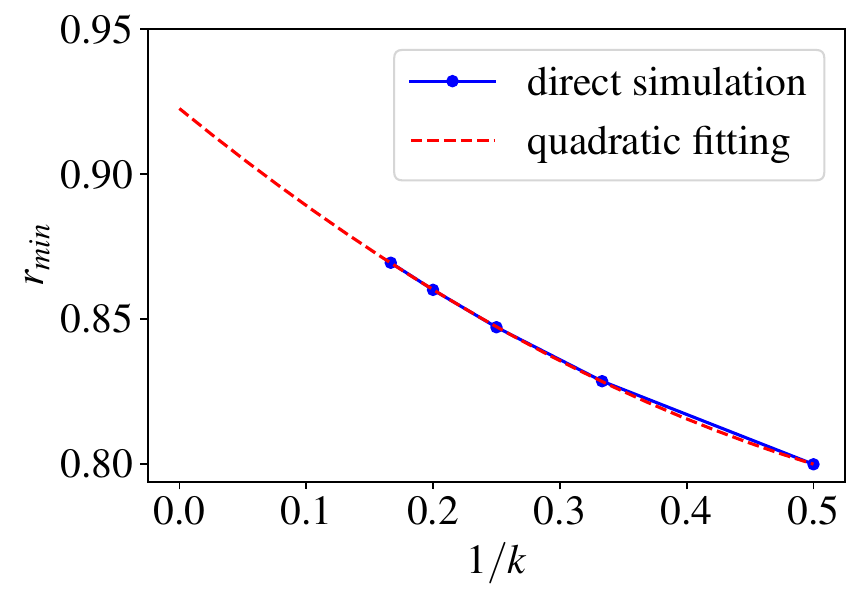}
    \caption{Minimum value of the order parameter $r_{\rm min} = \min_{\theta_1, \theta_2}(r)$ vs $1/k$, with $k$ the depth of the tree. A quadratic fit of the data is shown by the red dashed line, suggesting that $\lim_{k\to\infty}r_{\rm min} \simeq 0.92<1$.}
    \label{fig:min_order}
\end{figure}

To rule out the possibility of a finite-$k$ effect within a sharp phase, we study the behavior of the order parameter as a function of $k$. 
In Fig.~\ref{fig:min_order}, we plot the minimum of $r$ across the whole $\theta_{1,2}$ parameter space as a function of $k$. The data strongly suggests that $\lim_{k\to\infty} r_{min} < 1$; a quadratic fit to $1/k$ gives $r_{min} \simeq 0.92$ as $k\to\infty$. Overall, the numerical evidence backs up our analytical derivation of a nontrivial phase diagram at $p=1$, and the analytical phase boundaries derived in Eq.~\eqref{eq:SU2_critical_contour} appear consistent with numerical observations.

To summarize, we have found that the system is always in a fuzzy phase for $p < 1$, whereas for $p = 1$ it can be in either a sharp or fuzzy phase based on the choice of $\theta_1$ and $\theta_2$, with a spin-sharpening phase boundary determined by Eq.~\eqref{eq:SU2_critical_contour}. 
Considering the limitations of numerical simulation, it is still interesting to find methods that can reach larger system sizes and shed more light on the phase diagram and dynamics of this model. 
While this remains in general an open question for future work, we conclude this section by mentioning an efficient approach to study certain properties of the trajectory ensemble of very large trees.

The idea, already widely studied in the literature on measurement-induced phase transitions \cite{Bao_Theory_2020,Bao_2021}, is to modify the probability distribution over measurement outcomes in order to improve the analytical or numerical tractability of the problem. In particular, we take $\tilde{p}_n(\mathbf m) = p(\mathbf m)^n / \sum_{\mathbf m'} p(\mathbf m')^n$ with $n$ an integer. This includes {\it forced measurements}, where all trajectories are taken as equally likely ($n=0$), as well as the original Born-rule measurements ($n=1$) and other modified distributions ($n\geq 2$) that over-weight the more likely trajectories.
In Appendix~\ref{append:fake_measurement}, we numerically study these cases. Unlike the real measurement dynamics ($n=1$), these cases can be efficiently simulated: by the pool method for $n=0$, and by a small-sized recursion relation for $n\geq 2$. 
Again sticking to $p=1$, we find that forced measurements ($n=0$) yield only a sharp phase, while the $n\geq 2$ modified distributions yield an interesting phase diagram comprising both sharp and fuzzy phases. 
Interestingly, the $n\geq 2$ phase boundaries exhibit a strong asymmetry in $\theta_i \leftrightarrow \pi-\theta_i$, unlike the symmetric contour derived in Eq.~\eqref{eq:SU2_critical_contour} for the real measurement ($n=1$) case. Again in Appendix~\ref{append:fake_measurement} we show that an asymmetry with the same pattern is seen in the value of the order parameter $r$ within the $n=1$ fuzzy phase, confirming that some properties of the trajectory distribution are indeed changing at this new phase boundary. Understanding this phenomenon analytically is an interesting question that we leave for future work.

\section{Discussion}
\label{sec:concl}
In this work we addressed measurement-induced purification and charge sharpening transitions in dynamical quantum trees with abelian ($U(1)$) and non-abelian ($SU(2)$) symmetries. The recursive tree structure allowed us to locate critical points and identify critical behavior analytically in some cases, and using efficient numerical techniques in others, and thus to address questions that are inaccessible in other geometries: in particular, in the $U(1)$ case we showed that the purification and sharpening transitions immediately split when one introduces charge-neutral degrees of freedom; in the $SU(2)$ case we were able to identify a phase boundary between sharp and fuzzy phases in a model with maximal measurement rate. 

A particularly counterintuitive result in the $U(1)$ case is that the extent of the sharp phase is a nonmonotonic function of the dimension $d$ of neutral degrees of freedom. The sharpening transition at $d = \infty$ can be interpreted as a threshold for a suboptimal charge-learning algorithm for the $d = 1$ case~\cite{learnability_Fergus}, therefore $p_\# (d = \infty) \geq p_\# (d = 1)$. However, our numerical results unambiguously show that $p_\#(d = 2, 3) \leq p_\#(d = 1)$. Neutral degrees of freedom suppress quantum fluctuations in charge transport as well as fluctuations among the unitary gates; it seems that the former effect dominates when $d$ is small and the latter when $d$ is large. It would be interesting to understand whether this effect generalizes to other geometries. 

In the $SU(2)$ case, working with dynamical trees allowed us to find the first nontrivial analytical results on the phase diagram. However, much remains to be understood in this case: the sharp phase is extremely delicate, existing only in the limit where all qubits participate in a measurement at every time step. Whether this conclusion is specific to the tree geometry we chose, or occurs more generally, is an interesting open question. The nature of the critical point for $SU(2)$ symmetry is another question that will require a nontrivial extension of our methods to address. Finally, it is natural to ask whether there can be phase transitions inside the fuzzy or mixed phase, corresponding to the numerically observed transitions~\cite{majidy_critical_2023} between critical phases that sharpen/purify with different powers of system size. Our results based on alternative measurement outcome distributions (Appendix~\ref{append:fake_measurement}) are evidence of additional structure in the ensembles of trajectories, whose nature and consequences for dynamics remain to be understood.

Beyond these theoretical questions (and related ones for symmetry groups beyond those considered here), it would be interesting to explore these tree-like geometries in near-term experiments: techniques based on qubit reuse can in principle be used to study the expansion process on trees in a scalable manner~\cite{PRXQuantum.4.030334}.

\acknowledgements 
X.F. thanks Adam Nahum and Bernard Derrida for the useful discussion and suggestions. S.G. thanks Utkarsh Agrawal, David Huse, Ewan McCulloch, Drew Potter, Hans Singh, and Romain Vasseur for helpful discussions. This material is based upon work supported by the U.S. Department of Energy, Office of Science, National Quantum Information Science Research Centers, Co-design Center for Quantum Advantage (C2QA) under contract number DE-SC0012704 (theoretical analysis performed by S.G.).
Numerical simulations were performed in part on HPC resources provided by the Texas Advanced Computing Center (TACC) at the University of Texas at Austin.
\bibliographystyle{quantum}
\bibliography{sharpening_transition}

\appendix

\section{Recursive structure of dynamical trees with $U(1)$-symmetry}
\label{append:recur_struct_U1}
In Sec.~\ref{sec:U1_quant} and Sec.~\ref{sec:U1_class} we mention that the dynamics of $U(1)$-symmetric trees can be studied recursively. In this appendix, we give a detailed discussion of this fact.

\subsection{$d=1$}
First, we focus on the collapse tree (including the density matrix and outcome probability distribution) for $d=1$. A $k$-layer quantum tree can be viewed as a tree-like tensor network acting on an initial state. This tensor defines a mapping from a $2^k$-dimensional Hilbert space to a single qubit Hilbert space, which is the top qubit of the tree. We label the tensor represented by this tree as $T(\{U\},\mathbf{m})$. Here $\{U\}$ represents the set of unitary gates in the tree and $\mathbf{m}$ represents the locations of measurements and their corresponding outcomes. To simplify the notation, we ignore $\{U\}$ and just keep $T(\mathbf{m})$ as the argument of the tensor. Suppose the initial state is $\rho_0$, then the density matrix of the top qubit can be written as
\begin{equation}
    \rho = \frac{T(\mathbf{m})\rho_0 T^{\dagger}(\mathbf{m})}{\tr (T(\mathbf{m})\rho_0T^{\dagger}(\mathbf{m}))},
\end{equation}
and the probability of getting outcome $\mathbf{m}$ is
\begin{equation}
    p(\mathbf{m}) = (1-p)^{2^k-1-N_{\rm mid}}p^{N_{\rm mid}}\tr(T(\mathbf{m})\rho_0T^{\dagger}(\mathbf{m}))\label{eq:prob_total_U1}.
\end{equation}
Here $N_{\rm mid}$ represents the number of mid-circuit measurements in the tree. For the dynamics we are interested in, the initial state can always be the tensor product of local qubit density matrix, either mixed or pure. Our goal is to get a recursive construction of $T(\mathbf{m})$ (density matrix) and $p(\mathbf{m})$ (probability).

The recursive construction of the density matrix is easy to define. Considering the top node of the tree, there are two subtrees connecting to it. Since each subtree provides a single qubit density matrix as input, we can label the density matrices of the two subtrees by $\rho^1$ and 
$\rho^2$, which satisfy
\begin{equation}
    \rho^{i} = \frac{T^{i}(\mathbf{m}^i)\rho_0^{i} T^{i\,\dagger}(\mathbf{m}^i)}{\tr (T^i(\mathbf{m}^i)\rho_0^{i}T^{i\,\dagger}(\mathbf{m}^i))},\quad i=1,2.
\end{equation}
Here $T^i$ is the tensor of the subtree $i$. As mentioned above, the initial state is always separable, this allows us to write $\rho_0=\rho_0^{1}\otimes \rho_0^{2}$. We label the tensor of the single node by $t$, then we get $T=t(T^1\otimes T^2)$ and 
\begin{equation}
    \rho = \frac{t(\rho^1\otimes\rho^2)t^{\dagger}}{\tr (t(\rho^1\otimes\rho^2)t^{\dagger})}.
\end{equation}
Tensor $t$ is just the product of entangling gates and projectors in the node. Then we can get $\rho$ from $\rho_1$ and $\rho_2$ by the following steps: 
(i) we first construct the tensor product $\rho^1\otimes\rho^2$, 
(ii) we apply a random two-qubit $U(1)$-symmetric gate $U$ to get $\rho'=U(\rho^1\otimes\rho^2)U^{\dagger}$, 
(iii) we measure the second qubit along the $z$ direction with outcome $\sigma'$ and discard it from the system\footnote{This qubit can be randomly chosen since we have the statistical invariance under exchange of the two input qubits.}, 
(iv) we measure the remaining qubit along the $z$ direction with probability $p$ and we label the outcome by $\sigma$. 

In addition to the recursive construction of the density matrix, we also need to construct the probability distribution over measurement outcomes. Plugging $T= t(T^1\otimes T^2)$ into Eq.~\eqref{eq:prob_total_U1}, we get
\begin{align}
    p(\mathbf{m})=(1-p)^{2^k-1-N_{\rm mid}}p^{N_{\rm mid}}\tr\left[t(\rho^1\otimes\rho^2)t^{\dagger}\right]\tr\rho^1\tr\rho^2.
\end{align}
Since $N_{\rm mid}=N^1_{\rm mid}+N^2_{\rm mid}+\delta$ with $\delta=0,1$ representing the number of mid-circuit measurements in the node, we get
\begin{align}
    p(\mathbf{m})=(1-p)^{1-\delta}p^{\delta}\tr[t(\rho^1\otimes\rho^2)t^{\dagger}]p(\mathbf{m}^1)p(\mathbf{m}^2).
\end{align}
Here we have used Eq.~\eqref{eq:prob_total_U1} to obtain $p(\mathbf{m}^1)$ and $p(\mathbf{m}^2)$. The remaining factor can be understood as the conditional probability distribution over measurement outcomes inside the node, given the input states $\rho^1$ and $\rho^2$. To see this, when the input density matrices are fixed, a specific measurement outcome set $\{\sigma',\sigma\}$ has probability 
\begin{equation}
    p(\sigma',\sigma) = p\tr \left[ P_{\sigma'}U(\rho_1\otimes\rho_2)U^{\dagger}P_{\sigma'}\right]\label{eq:prob_double},
\end{equation}
while when we have only a single outcome $\sigma'$, the probability is
\begin{equation}
    p(\sigma') = (1-p)\tr \left[ P_{\sigma'}U(\rho_1\otimes\rho_2)U^{\dagger}P_{\sigma'}\right]\label{eq:prob_single}.
\end{equation}
 Introducing $\mathbf{m}'$ to label the number of mid-circuit measurement and their outcomes, we get
\begin{equation}
    p(\mathbf{m}'|\rho_1,\rho_2)=p^{\delta}(1-p)^{1-\delta}\tr[t(\rho_1\otimes\rho_2)t^{\dagger}].
\end{equation}
This immediately gives us 
\begin{align}
p(\mathbf{m})&=p(\mathbf{m}'|\mathbf{m}^1,\mathbf{m}^2)p(\mathbf{m}^1)p(\mathbf{m}^2).
\end{align}
Here we have omitted the dependence on random gates from our notation and use only $\mathbf{m}^1$ and $\mathbf{m}^2$ to label the density matrices of subtrees $1$ and $2$. 
This shows that the distribution of density matrices of the overall tree can be obtained recursively from the density matrices of the two subtrees based on the steps described above. By iterating this construction from the top to the bottom, we find that the whole tree can be constructed recursively.

From the above discussion, we see that the qubit tree can be constructed node by node. This ensures the efficient simulation of dynamics by the so called \textit{pool method} \cite{Miller_Weak-disorder_1994,Monthus_Anderson_2009,Garcia-Mata_Scaling_2017}, a way to get the final distribution recursively. 

\subsection{$d=\infty$}
In the large $d$ limit, the problem of charge sharpening transition can still be studied recursively. As mentioned in the main text, the problem becomes a classical random walk with constrains from the measurement outcomes. For each collapse tree, we define a vector $\mathbf p$ containing the probabilities of the top qubit to have charge $0$ or $1$. From the transition matrix of the random walk, we can still define a tensor $T(\mathbf{m})$ such that the unnormalized probability vector on the top is
\begin{equation}
    \mathbf{p} = T(\mathbf{m})\mathbf{p}_0.
\end{equation}
Here $\mathbf{p}_0 =(1/2,1/2)^{T\,\otimes 2^k}$ is the uniform distribution over all charge configurations of $2^k$ qubits. The probability of measurement outcome set $\mathbf{m}$ is simply
\begin{equation}
    p(\mathbf{m}) = (1-p)^{2^k-1-N_{\rm mid}}p^{N_{\rm mid}}\sum_{i=0,1} \mathbf{p}_i.
\end{equation}
The summation in above equation goes over the two components of $\mathbf{p}$, and it gives the probability of this specific measurement outcome set when the locations of measurements are fixed in the tree.
Similar to the $d=1$ case, $T(\mathbf{m})$ can be written as $t(T^1(\mathbf{m}^1)\otimes T^2(\mathbf{m}^2))$ with $T^1$ and $T^2$ corresponding to the two subtrees. Tensor $t$ becomes the product of matrix
\begin{equation}
    V =
    \begin{pmatrix}
        1&0&0&0\\
        0&1/2&1/2&0\\
        0&1/2&1/2&0\\
        0&0&0&1
    \end{pmatrix}.
\end{equation}
and measurement projectors. At the same time, suppose the two probability vectors of the two subtrees are $\mathbf{p}^1$ and $\mathbf{p}^2$, then the probability to get measurement outcome $\mathbf{m}'$ in the node is
\begin{equation}
    p(\mathbf{m}'|\mathbf{p}^1,\mathbf{p}^2)=(1-p)^{1-\delta}p^{\delta}\frac{\sum_i\left(t(\mathbf{p}^1\otimes\mathbf{p}^2)\right)_i}{\sum_i (\mathbf{p}^1)_i\sum_i (\mathbf{p}^2)_i},
\end{equation}
where $\delta$ is again the number of measurement in the node. 
Then the probability of $\mathbf{m}$ can be re-written as
\begin{align}
    p(\mathbf{m})=&(1-p)^{1-\delta}p^{\delta}\frac{\sum_i\left(t(\mathbf{p}^1\otimes\mathbf{p}^2)\right)_i}{\sum_i (\mathbf{p}^1)_i\sum_i (\mathbf{p}^2)_i}
    p(\mathbf{m}^1)p(\mathbf{m}^2)\\
    =&p(\mathbf{m}'|\mathbf{m}^1,\mathbf{m}^2)p(\mathbf{m}^1)p(\mathbf{m}^2).
\end{align}
These conditions ensure that the dynamics of charge sharpening can still be constructed node by node, and guarantee that we can use the pool method to efficiently simulate the dynamics.

So far, our discussion about the $U(1)$ tree's recursive construction is quite straightforward due to the fact that we have only real (i.e., Born rule) measurements in the whole collapse process. In Appendix~\ref{append:SU2_recur}, we are going to see a case where the dynamics cannot be simulated node by node, making the pool method inapplicable.

\section{Theoretical approach to phase transitions in the $U(1)$-symmetric quantum tree }
\label{append:lin_recur_U1}
In this section, we give a detailed theoretical derivation of the phase transitions in collapse trees with $U(1)$ symmetry.

\subsection{$d=1$}
As discussed in Appendix~\ref{append:recur_struct_U1}, a realization of our $U(1)$ quantum tree circuit can be represented by a tensor $T(\mathbf{m})$. The pair $(\mathcal{Z},s)$ of the whole tree can be obtained from spectrum of $\rho=T(\mathbf{m})\rho_0T^{\dagger}(\mathbf{m})/\tr(T(\mathbf{m})\rho_0T^{\dagger}(\mathbf{m}))$ with $\rho_0$ as the initial state. The tensor can be recursively obtained as $T(\mathbf{m})= t(T^1(\mathbf{m})\otimes T^2(\mathbf{m}))$.  For the tensor $t$ of a single node, it contains a random two-qubit $U(1)$-symmetric gate $U$ and a projective measurement with outcome $\sigma'$ to discard one qubit. There can be an extra projective measurement  on the remaining qubit with probability $p$. We label this measurement outcome by $\sigma$. 
Generally, $(\mathcal{Z},s)$ should be some nonlinear function of $(\mathcal{Z}_1,s_1)$, $(\mathcal{Z}_2,s_2)$ and $t$. Moreover, the probabilities of measurement outcomes should be determined based on the Born rule, which is also a nonlinear function of $\mathcal{Z}$. 

The nonlinearity of these recursive relations makes a solution of the problem quite challenging. 
However, in the pure/sharp phase, we have $\mathcal{Z}^{typ}\to 0$ when $k\to\infty$. Previous work~\cite{feng_measurement_2023} shows that the dynamics of $\ln \mathcal{Z}_k^{typ}$ is only controlled by the leading-order terms $\mathcal{Z}_1$, $\mathcal{Z}_2)$ in the recursive relation for $\mathcal{Z}$, along with the constant term in the conditional probability $p(\mathbf{m}'|\rho_1,\rho_2)$. All higher order corrections in $\mathcal{Z}_{1,2}$ can be neglected, as they are asymptotically smaller than the leading ones as one approaches the $\mathcal{Z}=0$ fixed point. This is still true in our problem. One thing that is different is the need to also keep track of the basis information $s$, so our recursive relations also depend on $s_1$ and $s_2$.

We first discuss the linearized recursive relation for the maximally mixed initial condition. Considering a specific node, when the remaining qubit is measured with outcome $\sigma$, we know that 
\begin{equation}
    \mathcal{Z} = 0,\quad s = \sigma
\end{equation}
independently of all the input data $\mathcal{Z}_{1,2}$, $s_{1,2}$ (note this is an exact statement).

When the node's outgoing qubit is not measured (this happens with probability $(1-p)$), the recursive relation defining $(\mathcal{Z},s)$ involves nonlinear functions of $\mathcal{Z}_{1,2}$, $\sigma'$ and $s_{1,2}$. Previous work~\cite{nahum_measurement_2021,feng_measurement_2023} shows that the phase transition is determined by expanding these function and keeping only terms linear in $\mathcal{Z}_{1,2}$. All higher order terms in the Taylor expansion don't change the critical point. Keeping only the leading terms in $\mathcal{Z}_{1,2}$, we have the recursive relation of the pair $(\mathcal{Z},s)$ in terms of $\mathcal{Z}_{1,2}$, $\sigma'$ and $s_{1,2}$:
\begin{itemize}
    \item if $s_1\ne s_2$, then 
    \begin{equation}
        \mathcal{Z} \approx \left|U^{1-\sigma',\sigma'}_{s_1,s_2}\right|^{-2} \mathcal{Z}_{1+\sigma'}, \quad s=1-\sigma';
    \end{equation}
    \item if $\sigma' = s_1=s_2$, then 
    \begin{equation}
        \mathcal{Z} \approx \left|U^{1-\sigma',\sigma'}_{1-\sigma',\sigma'}\right|^2 \mathcal{Z}_{1} + \left|U^{1-\sigma',\sigma'}_{\sigma',1-\sigma'}\right|^2\mathcal{Z}_2,
        \quad s=\sigma'.
        \label{eq:linear_recur_U1d1}
    \end{equation}
\end{itemize}
The above results cannot fully determine the statistics of $\mathcal{Z}$. What is missing is the conditional probability to have measurement outcome $\sigma'$ with input $\mathcal{Z}_{1,2}$ and $s_{1,2}$. This is generally hard to solve. However, it was shown \cite{feng_measurement_2023} that to solve the phase transition, we only need to keep the terms independent of $\mathcal{Z}_{1,2}$ in the Taylor expansion of this conditional probability. The constant term $p^0$ of the conditional probability $p(\mathbf{m}'|\mathcal{Z}_1,s_1,\mathcal{Z}_2,s_2)$ (i.e., the part independent of $\mathcal{Z}_{1,2}$) satisfies
\begin{equation}
    p^0 = \left\{ 
    \begin{aligned} 
    & (1-p)\left|U^{1-\sigma',\sigma'}_{s_1,s_2}\right|^2 & \text{if } s_1\ne s_2 \\
    &(1-p) & \text{if } \sigma'=s_1=s_2.
    \end{aligned} \right. \label{eq:prob_constant_U1d1}
\end{equation}
Notice that in Eq.~\eqref{eq:linear_recur_U1d1} and Eq.~\eqref{eq:prob_constant_U1d1}, when $s_1=s_2$, we only consider the case with $\sigma'=s_1=s_2$. In principle, there should still be nonzero probability for $\sigma'\ne s_{1}$ as long as $\mathcal{Z}_{1}(\mathcal{Z}_2)\ne 0$. However, this case gives $p(\mathbf{m}'|\mathcal{Z}_1,s_1,\mathcal{Z}_2,s_2)$ linear in $\mathcal{Z}_{1}(\mathcal{Z}_2)$ such that $p^0=0$. For the purpose of getting the critical point, we can safely ignore this case \cite{feng_measurement_2023}. 

To better understand the dynamics with large enough $k$, we introduce the generating function $G_k(x)=\sum_{\mathcal{Z}_k}\sum_{s_k} p(\mathcal{Z}_k,s_k)\exp(-e^{-x}\mathcal{Z}_k)$. Here the average goes over all possibilities of the pair $(\mathcal{Z}_k,s_k)$. When the initial state is invariant under global charge conjugation, we can see that $p(s=0,\mathcal{Z})=p(s=1,\mathcal{Z})$. This gives $p(s,\mathcal{Z})=p(\mathcal{Z})/2$. Then we can rewrite $G_k(x)=\langle \exp(-e^{-x}\mathcal{Z}_k)\rangle$ with the average taken over only the distribution of $\mathcal{Z}$ value. Generally, $G_k(x)$ can be viewed as a moving wavefront located at $\ln \mathcal{Z}^{typ}_k$. At late time, this wavefront can be described by a velocity $v_p$ \cite{Derrida_Polymers_1988,nahum_measurement_2021,feng_measurement_2023},
\begin{equation}
    G_k(x)= G^{(\lambda)}(x-v_p(\lambda)t),
\end{equation}
with $\lambda$ as a parameter characterizing the family of possible moving wave solutions. The actual physical one is that with the minimal velocity \cite{Derrida_Polymers_1988}. By solving the velocity $v_p$, we can determine the phase transition. In the pure/sharp phase, we have $v_p<0$ since $\mathcal{Z}^{typ}_k$ decays to $0$ with increase of $k$; and the ballistic velocity $v_p$ vanishes at the critical point \cite{nahum_measurement_2021}. Now our task is to analytically get $v_p$. This can be done by considering the recursive relation of $G_k(x)$.
Plugging Eq.~\eqref{eq:linear_recur_U1d1} and Eq.~\eqref{eq:prob_constant_U1d1} into the definition of $G_k(x)$, we get,
\begin{widetext}
\begin{align}
    G_{k+1}(x)&= \sum_{s_1,s_2}\sum_{\sigma'}\sum_{\mathcal{Z}_1,\mathcal{Z}_2} \exp(-e^{-x}\mathcal{Z}(s_1,s_2,\sigma',\mathcal{Z}_1,\mathcal{Z}_2)) (1-p)p(\sigma'|\mathcal{Z}_1,s_1,\mathcal{Z}_2,s_2)p(s_1,\mathcal{Z}_1)p(s_2,\mathcal{Z}_2)+p\nonumber\\
    &=\frac{1-p}{4}\sum_{s_1,s_2}\sum_{\sigma'} \sum_{\mathcal{Z}_1,\mathcal{Z}_2}\exp(-e^{-x}\mathcal{Z}(s_1,s_2,\sigma',\mathcal{Z}_1,\mathcal{Z}_2)) p(\sigma'|\mathcal{Z}_1,s_1,\mathcal{Z}_2,s_2)p(\mathcal{Z}_1)p(\mathcal{Z}_2)+p\nonumber\\
    &\approx \frac{1-p}{4}\sum_{s_1+s_2=1}\sum_{\sigma'}\left\langle A_{s_1s_2}(\sigma')G_k(x+\ln A_{s_1s_2}(\sigma'))\right\rangle_U+\frac{1-p}{4}\sum_{s_1}\left\langle G_k(x-\ln B_{s_1})G_k(x-\ln C_{s_1})\right\rangle_U+p\label{eq:recur_G_U1d1},
\end{align}
\end{widetext}
with coefficients
\begin{equation}
\begin{aligned}
    A_{s_1, s_2}(\sigma') & =\left|U_{s_1,s_2}^{1-\sigma',\sigma'}\right|^2, \\
    B_{s_1}& =\left|U^{1-s_1,s_1}_{1-s_1,s_1}\right|^2,\\
    C_{s_1}& =\left|U_{s_1, 1-s_1}^{1-s_1, s_1}\right|^2.
\end{aligned}
\end{equation}
The last term $p$ in Eq.~\eqref{eq:recur_G_U1d1} comes from the fact that when the remaining qubit of the node is measured, then $\mathcal{Z}=0$. The average $\langle\dots\rangle$ is taken over all random two-qubit gates which maintain the $U(1)$ symmetry. In the limit of large argument of $G^{(\lambda)}(u)$, the exponential decay of $G^{(\lambda)}$ is controlled by $\lambda$ \cite{Derrida_Polymers_1988}, 
\begin{equation}
    G^{(\lambda)}(u)\sim 1-e^{-\lambda u}.
\end{equation}
Combining with Eq.~\eqref{eq:recur_G_U1d1}, we get
\begin{align}
    v_p(\lambda) =& \frac{1}{\lambda}\ln\left[ \frac{1-p}{4}\sum_{s_1}\sum_{\sigma'}\langle A_{s_1, 1-s_1}^{1-\lambda}(\sigma')\rangle\right.\nonumber\\
    &\quad\quad\left.+\sum_{s_1}\left(\langle B_{s_1}^{\lambda}\rangle+\langle C_{s_1}^{\lambda}\rangle\right)\right].
\end{align}
Considering the fact that $U$ is a random gate with $U(1)$ symmetry, we get
\begin{widetext}
\begin{align}
v_p(\lambda) = \frac{1}{\lambda}&\ln\left[\frac{1-p}{2}\left(\left\langle\left|u_1\right|^{2(1-\lambda)}+\left|u_2\right|^{2(1-\lambda)}+\left|u_1\right|^{2\lambda}+\left|u_2\right|^{2\lambda}\right\rangle_u\right)\right].
\end{align}
\end{widetext}
Here $u=(u_1,u_2)$ is a uniformly distributed normalized complex vector. For fixed $p$, there is a $\lambda^{\ast}$ which gives the minimal velocity $\frac{\partial v_p(\lambda)}{\partial \lambda}|_{\lambda^{\ast}}=0$. However, the initial condition gives $G_0(x)\sim 1-\mathcal{Z}_0e^{-x}$ when $x$ is large. This requires that the actual $\lambda\le 1$ \cite{Derrida_Polymers_1988}. Then at the critical point $p_c$, we have
\begin{align}
    v_{p_c} =0,\quad\lambda = \min(\lambda^{\ast},1).
\end{align}
Taking the derivative of the velocity over $\lambda$, we get
\begin{widetext}
\begin{align}
    &-\frac{1}{\lambda^{\ast 2}}\ln\left[(1-p)\left(\left\langle\left|u_1\right|^{2(1-\lambda^{\ast})}+\left|u_1\right|^{2\lambda^{\ast}}\right\rangle_u\right)\right]\nonumber+\frac{2\left\langle \ln\left|u_1\right|\left(\left|u_1\right|^{2\lambda^{\ast}}-\left|u_1\right|^{2(1-\lambda^{\ast})}\right)\right\rangle_u}{\lambda^{\ast}\left\langle\left|u_1\right|^{2(1-\lambda^{\ast})}+\left|u_1\right|^{2\lambda^{\ast}}\right\rangle_u}=0.
\end{align}
\end{widetext}
At the critical point, $v_p=0$, this gives
\begin{align}
    &\left\langle \ln\left|u_1\right|\left(\left|u_1\right|^{2\lambda^{\ast}}-\left|u_1\right|^{2(1-\lambda^{\ast})}\right)\right\rangle_u=0.
\end{align}
We immediately get that $\lambda^{\ast}=1/2$. So at the critical point, we have $\lambda=1/2$. This requires that the critical point satisfies
\begin{equation}
    2(1-p)\left\langle\left|u_1\right|\right\rangle_u=1.
\end{equation}
Then we get the critical point,
\begin{equation}
    p_c = \frac{1}{4}.
\end{equation}

In previous work \cite{nahum_measurement_2021,feng_measurement_2023}, a complete theory about the scaling exponent was established, which also works for our recursive relation of $G_k(x)$ in Eq.~\eqref{eq:recur_G_U1d1}. It was proved that $\lambda$ determines the universality class of the phase transition. In our case, $\lambda=1/2$, so we are in the glass class and we get Eq.~\eqref{eq:U1_scale}.

Now we turn to the ``sharpening'' setup, with a pure initial state which is the superposition of all bitstrings with equal weight. The main difference is that the two input states of each node are now pure states with the density matrices
\begin{align}
    \rho = &(1-\mathcal{Z})\ket{1-s}\bra{1-s}+\mathcal{Z}\ket{s}\bra{s}\nonumber\\
    &\quad+\sqrt{\mathcal{Z}(1-\mathcal{Z})}X,
\end{align}
with $X$ the Pauli-$x$ matrix.
We see that now there are extra off-diagonal matrix elements in $\rho$. Our dynamics guarantees that this form is maintained for the output density matrix (a phase factor in the off-diagonal entries can be absorbed into the unitary gates). This changes the recursive relation. We find that when the outgoing qubit is measured (probability $p$), we still have
\begin{equation}
    \mathcal{Z} = 0,\quad s= \sigma,
\end{equation}
with $\sigma$ the measurement outcome. But when the the outgoing qubit is not measured (probability $1-p$), the recursive relation changes. We find that:
\begin{itemize}
    \item if $s_1\ne s_2$, then 
    \begin{align}
        \mathcal{Z} \approx & \left|U^{1-\sigma',\sigma'}_{s_1,s_2}\right|^{-2} \mathcal{Z}_{1+\sigma'}, \quad s=1-\sigma';
    \end{align}
    \item if $\sigma'=s_1=s_2$, then 
    \begin{align}
        \mathcal{Z} \approx &\left|U^{1-\sigma',\sigma'}_{1-\sigma',\sigma'} \sqrt{\mathcal{Z}_{1}}+U^{1-\sigma',\sigma'}_{\sigma',1-\sigma'}\sqrt{\mathcal{Z}_2}\right|^2,\quad s=\sigma'.\label{eq:nonlinear_recur_U1d1}
    \end{align}
\end{itemize}
In addition we have
\begin{equation}
    p^0 = \left\{
    \begin{aligned} 
    & (1-p)\left|U^{1-\sigma', \sigma'}_{s_1, s_2}\right|^2 & \text{if} \ s_1\ne s_2 \\
    &(1-p) & \text{if} \ \sigma'=s_1=s_2.
    \end{aligned} \right.
\end{equation}
Again when $s_1=s_2$, we ignore the case of $\sigma'\ne s_{1}$ since in this case the leading term in the conditional probability is $O(\mathcal{Z}_{1,2})$. 

The difference with the previous case is that when $s_1=s_2$, there is a nonlinear term $\sqrt{\mathcal{Z}_1\mathcal{Z}_2}$ in the leading order of the recursive relation for $\mathcal{Z}$. This forbids us from simply using the previous method to study the transition. Following the argument in the main text, we conjecture that this case has the same critical point as the case of a maximally mixed initial state. It would be interesting to develop an analytical solution to this phase transition in future work.

\subsection{Large-$d$ limit}
In the limit $d\to\infty$, the dynamics becomes a classical problem. 
For the purification phase transition, we have a percolation problem on the tree. The minimal domain wall cuts at most one link. Suppose the probability that the tree is connected (minimal domain wall length of 1) is $p_{\rm connect}$; then we have
\begin{equation}
    1-p_{\rm connect}=p+(1-p)(1-p_{\rm connect})^2.
\end{equation}
This condition gives $p_c=1/2$. When $p>p_c$, the tree becomes disconnected, and we are in the pure phase.

For charge sharpening transition, it is characterized by the uncertainty about charge on the unmeasured links of the domain wall. This is equivalent to asking about the uncertainty of the charge of the top qubit. Given an arbitrary subtree, we introduce a probability vector $(p(0),p(1))^T$ to represent the probability of the charge being $0$ or $1$ at its top qubit. By defining the $\mathcal{Z}$ value as the smaller of the two probabilities and $\ket{s}$ as the state with larger probability, we can again consider the recursive relation of $(\mathcal{Z},s)$. Suppose the two input subtrees are labeled by $(\mathcal{Z}_1,s_1)$ and $(\mathcal{Z}_2,s_2)$, we have the corresponding probability vector
\begin{equation}
 \begin{pmatrix}
        (1-\mathcal{Z}_{i})s_{i}+\mathcal{Z}_{i}(1-s_{i})\\\\
        \mathcal{Z}_{i}s_{i}+(1-\mathcal{Z}_{i})(1-s_{i})
    \end{pmatrix},
\end{equation}
with $i=1,2$. The tensor product of these two vectors defines the input of this node. The output probability vector is obtained by acting tensor $t$ on the input. When the outgoing qubit is measured with outcome $\sigma$, we have
\begin{equation}
    \mathcal{Z} = 0,\quad s= \sigma.
\end{equation}
When the outgoing qubit is not measured, our derivation shows
\begin{equation}
    \left\{ 
    \begin{aligned}
    \mathcal{Z} & \approx 2\mathcal{Z}_{1+\sigma'},
    \quad s =1-\sigma',
    \quad \text{if}\,\, s_1\ne s_2, \\
    \mathcal{Z} & \approx \frac{1}{2} \mathcal{Z}_{1}+\frac{1}{2}\mathcal{Z}_2,
    \quad s =\sigma',
    \quad \text{if}\,\,\sigma'=s_1=s_2.
    \end{aligned} \right. 
    \label{eq:linear_recur_U1dinf}
\end{equation} 
Here the coefficients are fixed since $V$ is a constant matrix. The constant piece of the measurement outcome probability becomes
\begin{equation}
    p^0 = \left\{ 
    \begin{aligned}
    &\frac{(1-p)}{2},\quad\text{if}\,\, s_1\ne s_2 \nonumber\\
    &(1-p),\quad \text{if}\,\, \sigma'=s_1=s_2.
    \end{aligned}\right.
    \label{eq:prob_constant_U1dinf}
\end{equation}

We see that the recursive relations for the charge sharpening transition with $d\to\infty$ have similar forms as those in the purification transition with $d=1$. Using the same methods, we get $\lambda =\frac{1}{2}$ and $p_{\#}=1-\frac{\sqrt{2}}{2}$. The result $\lambda=\frac{1}{2}$ guarantees that the scaling exponents are still those in Eq.~\eqref{eq:U1_scale}.

\section{$U(1)$-symmetric tree with $d=3$}
\label{appendix:U1_d3}
\begin{figure}
    \centering
    \includegraphics[width=0.9\columnwidth]{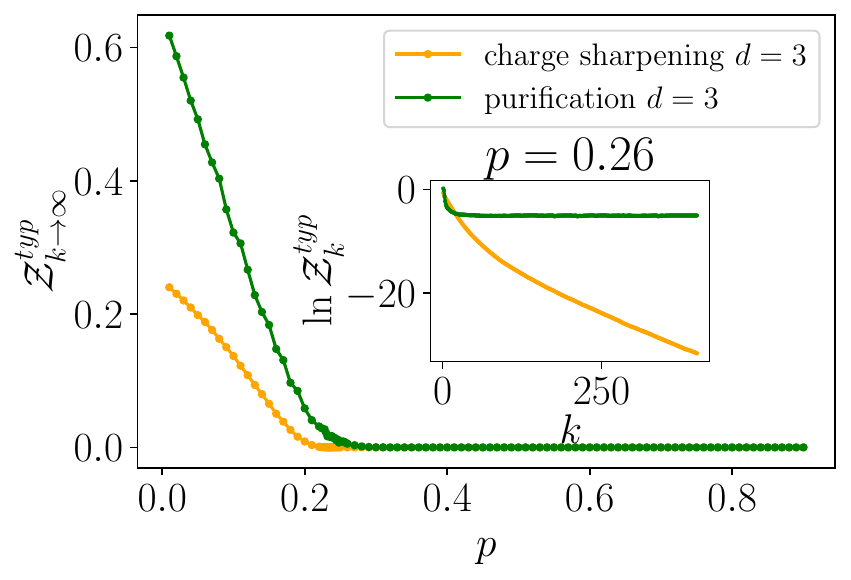}
    \caption{Numerical simulation of purification and charge sharpening transitions at $d=3$. The error bars are smaller or equal to the marker size. All data are obtained by the pool method with pool size equal to $10^6$. The inset shows dynamics of $\ln \mathcal{Z}^{typ}_k$ in both the purification process and charge sharpening process at $p=0.26$.}
    \label{fig:U1_d3}
\end{figure}

\begin{figure}
    \centering
    \includegraphics[width=0.9\columnwidth]{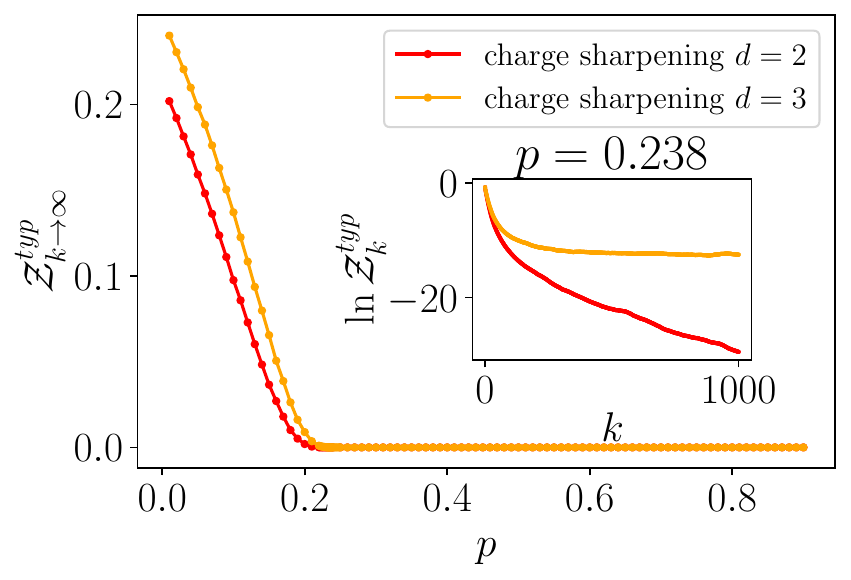}
    \caption{Comparison of charge sharpening transition at $d=2$ and $d=3$. The errorbars are smaller or equal to the marker size. All data are obtained by the pool method with pool size equal to $10^6$. The inset gives the curves of $\ln \mathcal{Z}_k^{typ}$ with $d=2$ and $d=3$ at $p=0.238$.}
    \label{fig:charge_sharp_d2_3}
\end{figure}

In the main text, we briefly mention the purification and charge sharpening transitions at $d=3$. Here we show numerical simulations of these transitions. Numerical results for the purification transition and charge sharpening transition are shown in Fig.~\ref{fig:U1_d3}. 
To better compare the critical points, we consider a specific probability $p=0.26$ at which we can clearly see that the curve of $\ln \mathcal{Z}_k^{typ}$ saturates in the purification process, while the one for the charge shaprening process decays to $-\infty$. This confirms that $p_{\#}<p_c$ at $d=3$. Furthermore, we compare the charge sharpening transitions with $d=2$ and $d=3$ in Fig.~\ref{fig:charge_sharp_d2_3}. From it we can conclude that $p_{\#}$ increases when $d$ increases from $2$ to $3$.

\section{Recursive construction of quantum tree with $SU(2)$ symmetry}
\label{append:SU2_recur}

\begin{table*}
\centering
\begin{adjustbox}{width=0.9\textwidth}
\renewcommand{\arraystretch}{2}
\begin{tabular}{|c|c|c|c|c|}
   \hline
   \diagbox{$m's$}{input}&$P_s\otimes P_s$&$P_s\otimes P_t$& $P_t\otimes P_s$& $P_t\otimes P_t$\\
   \hline
   $\mathds{1}$,$\mathds{1}$ &$\frac{1}{4}P_s$&$\frac{1}{4}P_t$&$\frac{1}{4}P_t$&$\frac{1}{2}P_t+\frac{3}{4}P_s$\\
   \hline
   $P_s$,$\mathds{1}$&$\frac{1}{32}(5+3c_{+})P_s$&$(\frac{3}{32}+\frac{c_1}{16}+\frac{c_2}{16}+\frac{c_{-}}{32})P_t$&$(\frac{3}{32}+\frac{c_1}{16}+\frac{c_2}{16}+\frac{c_{-}}{32})P_t$& $(\frac{15}{32}+\frac{9c_{+}}{32})P_s+(\frac{3}{16}+\frac{c_1}{8}+\frac{c_2}{8}+\frac{c_{-}}{16})P_t$\\
   \hline
   $P_t$,$\mathds{1}$&$\frac{9}{32}(1-c_{+})P_s$&$(\frac{7}{32}-\frac{3c_1}{16}+\frac{c_2}{16}-\frac{3c_{-}}{32})P_t$&$(\frac{7}{32}-\frac{3c_2}{16}+\frac{c_1}{16}-\frac{3c_{-}}{32})P_t$&$\frac{3}{32}(1-c_{+})P_s+(\frac{3}{16}-\frac{c_1}{8}-\frac{c_2}{8}+\frac{c_{-}}{16})P_t$\\
   \hline
   $\mathds{1}$,$P_s$&$(\frac{5}{32}+\frac{3}{32}c_{+})P_s$&$(\frac{3}{32}-\frac{c_1}{16}-\frac{c_2}{16}+\frac{c_{-}}{32})P_t$&$(\frac{3}{32}-\frac{c_1}{16}-\frac{c_2}{16}+\frac{c_{-}}{32})P_t$&$(\frac{15}{32}+\frac{9}{32}c_{+})P_s+(\frac{3}{16}-\frac{c_1}{8}-\frac{c_2}{8}+\frac{c_{-}}{16})P_t$\\
   \hline
   $\mathds{1}$,$P_t$& $\frac{9}{32}(1-c_{+})P_s$&$(\frac{7}{32}+\frac{3c_1}{16}-\frac{c_2}{16}-\frac{3}{32}c_{-})P_t$&$(\frac{7}{32}+\frac{3c_2}{16}-\frac{c_1}{16}-\frac{3}{32}c_{-})P_t$&$\frac{3}{32}(1-c_{+})P_s+(\frac{3}{16}+\frac{c_1}{8}+\frac{c_2}{8}+\frac{c_{-}}{16})P_t$\\
   \hline
   $P_s$,$P_s$&$(\frac{5}{32}+\frac{3}{32}c_{+})P_s$&$0$&$0$&$(\frac{15}{32}+\frac{9}{32}c_{+})P_s$\\
   \hline
   $P_s$,$P_t$ &$0$&$(\frac{3}{32}+\frac{c_1}{16}+\frac{c_2}{16}+\frac{c_{-}}{32})P_t$&$(\frac{3}{32}+\frac{c_1}{16}+\frac{c_2}{16}+\frac{c_{-}}{32})P_t$&$(\frac{3}{16}+\frac{c_1}{8}+\frac{c_2}{8}+\frac{c_{-}}{16})P_t$\\
   \hline
   $P_t$,$P_s$&$0$&$(\frac{3}{32}-\frac{c_1}{16}-\frac{c_2}{16}+\frac{c_{-})}{32})P_t$&$(\frac{3}{32}-\frac{c_1}{16}-\frac{c_2}{16}+\frac{c_{-}}{32})P_t$&$(\frac{3}{16}-\frac{c_1}{8}-\frac{c_2}{8}+\frac{c_{-}}{16})P_t$\\
   \hline   $P_t$,$P_t$&$\frac{9}{32}(1-c_{+})P_s$&$\frac{1}{8}(1-c_{-})P_t$&$\frac{1}{8}(1-c_{-})P_t$&$\frac{3}{32}(1-c_{+})P_s$\\
   \hline
\end{tabular}
\end{adjustbox}
\caption{Table of the node outputs $t(P_i\otimes P_j)t^{\dagger}$ with $i,j=s$ (singlet) or $t$ (triplet). The top row labels the node input $P_i\otimes P_j$, while the leftmost column labels the possible outcomes of measurement within the node: $\mathds{1}$ (no measurement), $P_s$ (measurement with singlet outcome), $P_t$ (measurement with triplet outcome). The first listed outcome corresponds to the inner pair of qubits while the second corresponds to the outer pair of qubits, see Fig.~\ref{fig:SU2_tree}. To simplify the notation, we introduce symbols $c_{i}=\cos(2\theta_{i})$ for $i=1,2$ and $c_{\pm}=\cos(2\theta_1\pm2\theta_2)$, with $\theta_1$ and $\theta_2$ the gate angle parameters in the node.}
\label{tab:output_SU2}
\end{table*}

In this section, we give a more detailed discussion about the recursive structure of our expansion tree with $SU(2)$ symmetry. As mentioned in Sec.~\ref{sec:spin_sharpening}, for a $k$-layer expansion tree, the density matrix of the two probe qubits takes the form
\begin{equation}
    \rho = \sigma P_s+\frac{\tau
    }{3}P_t,
\end{equation}
with $\sigma+\tau$ equal to the probability of getting measurement outocomes $\mathbf{m}$.
We are interested in learning how the distribution of $(\sigma,\tau)$ changes with $k$. More specifically, our task is to calculate
\begin{equation}
     r =\sum_{\mathbf{m}}\frac{\sigma(\mathbf{m})+\tau(\mathbf{m})}{\sum_{\mathbf{m}}(\sigma(\mathbf{m})+\tau(\mathbf{m}))}\frac{\sigma(\mathbf{m})^2+\tau(\mathbf{m})^2}{(\sigma(\mathbf{m})+\tau(\mathbf{m}))^2},
\end{equation}
which serves as an order parameter for spin sharpening. 
To obtain the recursive relation between density matrices, we first describe an arbitrary realization of the tree by a tensor $T(\mathbf{m})$, which establishes a map from the Hilbert space of two qubits to the Hilbert space of $2^{k_{\rm tot}+1}$ qubits with $k_{\rm tot}$ the total number of layers in the tree. The initial state of the system and probe qubits takes the form
\begin{equation}
    \ket{\psi_0}=\frac{1}{\sqrt{2}}(\ket{0}_1\ket{0}_{1'}+\ket{1}_1\ket{1}_{1'})\otimes \frac{1}{\sqrt{2}}(\ket{0}_2\ket{0}_{2'}+\ket{1}_2\ket{1}_{2'}),
\end{equation}
with subscripts $1,2$ representing the two initial system qubits and $1',2'$ representing the two probe qubits. The final state of the system and probe qubits together is
\begin{equation}
    \ket{\psi_f}=T_{12}(\mathbf{m})\ket{\psi_0}.
\end{equation}
Here we use the notation $T_{12}$ to clarify that $T$ only acts on the Hilbert space of two initial qubits 1, 2, while it acts trivially on the reference qubits $1'$, $2'$. From this we can immediately see that the final density matrix of the two probe qubits is
\begin{align}
    \rho = &\frac{\tr_{\rm sys} \left(T_{12}(\mathbf{m})\ket{\psi_0}\bra{\psi_0}T^{\dagger}_{12}(\mathbf{m})\right)}{\tr\left( T_{12}(\mathbf{m})\ket{\psi_0}\bra{\psi_0}T^{\dagger}_{12}(\mathbf{m})\right)}\nonumber\\
    =& \frac{\sum_{\lambda,\lambda'}\ket{\lambda}_{1'2'}\bra{\lambda'}_{12}T^{\dagger}_{12}(\mathbf{m})T_{12}(\mathbf{m})\ket{\lambda}_{12}\bra{\lambda'}_{1'2'}}{\tr\left(T^{\dagger}_{12}(\mathbf{m})T_{12}(\mathbf{m})\right)}\nonumber\\
    =& \frac{\left(T^{\dagger}(\mathbf{m})T(\mathbf{m})\right)^T}{\tr \left(T^{\dagger}(\mathbf{m})T(\mathbf{m})\right)}\label{eq:dm_probe}.
\end{align}
In the second equivalence $\lambda,\lambda'$ are in the basis $\ket{00},\ket{01},\ket{10}$ and $\ket{11}$. In the third equivalence we ignore the subscript $12$ and assume that when there is no subscript, the density matrix is defined in the Hilbert space of the probe qubits. 

As mentioned before, the density matrix Eq.~\eqref{eq:dm_probe} should takes the form $\sigma P_s+\frac{\tau}{3}P_t$ due to the $SU(2)$ symmetry. The probability to get $\mathbf{m}$ is
\begin{equation}
    p(\mathbf{m}) = \frac{1}{4}p^{N_{\rm mid}}(1-p)^{2^{k_{\rm tot}+1}-2-N_{\rm mid}}\tr(T^{\dagger}(\mathbf{m})T(\mathbf{m}))\label{eq:prob_m},
\end{equation}
with $N_{\rm mid}$ representing the total number of measurements in the tree. Considering the tree reversely from the top to the bottom, we can view it as connecting two subtrees with $k_{\rm tot}-1$ layers to a node. This is in fact mapping the expansion tree to its reverse collapse process. Defining $T'(\mathbf{m}')$, $T^{\prime\prime}(\mathbf{m}^{\prime\prime})$ and $t(\tilde{\mathbf{m}})$ to represent the two subtrees and the node repsectively, we get
\begin{equation}
    T(\mathbf{m})=\left(T'(\mathbf{m}')\otimes T^{\prime\prime}(\mathbf{m}^{\prime\prime})\right)t(\tilde{\mathbf{m}}).
\end{equation}
This enables us to re-write Eq.~\eqref{eq:dm_probe} as
\begin{align}
    \rho = \frac{\left(t^{\dagger}(\tilde{\mathbf{m}})\rho'\otimes\rho^{\prime\prime}t(\tilde{\mathbf{m}})\right)^T}{\tr\left(t^{\dagger}(\tilde{\mathbf{m}})\rho'\otimes \rho^{\prime\prime}t(\tilde{\mathbf{m}})\right)},
\end{align}
with
\begin{equation}
    \rho^{\prime(\prime\prime)} = \frac{T^{\prime(\prime\prime)\dagger}(\mathbf{m})T^{\prime(\prime\prime)}(\mathbf{m})}{\tr \left(T^{\prime(\prime\prime)\dagger}(\mathbf{m})T^{\prime(\prime\prime)}(\mathbf{m})\right)}.
\end{equation}

This shows that if we know the density matrices associated to the two subtrees, we can recursively construct the density matrix associated to the whole tree\footnote{In principle, the actual density matrix needs to be transposed. But for our problem the density matrix is real symmetric, so the transpose can be ignored.}. The above results give the normalized density matrix of the probe qubits. In principle, we can drop the denominator and consider the recursive relation of the unnormalized density matrix $\tilde{\rho}$
\begin{align}
    &\tilde{\rho}=\left(t^{\dagger}(\tilde{\mathbf{m}})\tilde{\rho}'\otimes\tilde{\rho}^{\prime\prime}t(\tilde{\mathbf{m}})\right)^T\nonumber,\\
    &\tilde{\rho}^{\prime(\prime\prime)} = T^{\prime(\prime\prime)\dagger}(\mathbf{m})T^{\prime(\prime\prime)}(\mathbf{m}).\label{eq:tilde_rho_recur}
\end{align}
Then we can directly get $\sigma$ and $\tau$ from the fact that $\tilde{\rho}=\sigma P_s+\tau P_t/3$. 

For the probability distribution over measurement outcomes, the situation is a little different. We can still write down a recursive relation,
\begin{align}
    p(\mathbf{m})&=4 p^{N(\tilde{\mathbf{m}})}(1-p)^{2-N(\tilde{\mathbf{m}})}\tr\left(t^{\dagger}(\tilde{\mathbf{m}})\rho'\otimes\rho^{\prime\prime}t(\tilde{\mathbf{m}})\right)\nonumber\\
    &\quad \times p(\mathbf{m}')p(\mathbf{m}^{\prime\prime})\nonumber\\
    &=4p(\tilde{\mathbf{m}}|\mathbf{m}',\mathbf{m}^{\prime\prime})p(\mathbf{m}')p(\mathbf{m}^{\prime\prime})\label{eq:prob_recur_SU2}.
\end{align}
However, crucially, the term $4p(\tilde{\mathbf{m}}|\mathbf{m}',\mathbf{m}^{\prime\prime})$ cannot be simply understood as a new conditional probability. Thus we cannot construct the distribution of measurement outcomes node by node, as in the $U(1)$ case. This prevents us from simulating the dynamics by the pool method. This is the main challenge in the study of the $SU(2)$ quantum tree. 

To better understand why this happens, we go back to our expansion process. In every node, we have a singlet pair as the input. When we want to consider the reverse collapse process, we notice that all singlet initial pairs now have to be considered as forced measurement with outcome to be singlet. This means that we need to \textit{post-select} trajectories which give singlet outcome on all these measurements. The post-selection changes the global distribution of measurement outcomes, which now cannot be constructed solely by considering the local probability distribution within a single node.

Now the main task is to obtain Eq.~\eqref{eq:sigma_tau_recur}. Considering a $k$-layer tree as connecting two subtrees to a single node, we want to get the parameters $(\sigma,\tau)$ of the whole tree from those of the two subtrees, $(\sigma_1,\tau_1)$ and $(\sigma_2,\tau_2)$. We notice that in Eq.~\eqref{eq:tilde_rho_recur}, the input of the node can be re-written as
\begin{equation}
    \sigma_1\sigma_2 P_s\otimes P_s+\frac{\sigma_1\tau_2}{3}P_s\otimes P_t+\frac{\tau_1\sigma_2}{3}P_t\otimes P_s+\frac{\tau_1\tau_2}{9}P_t\otimes P_t.
\end{equation}
This allows us to view $P_{i}\otimes P_{j}$, $i,j=s$ or $t$ as the basis. Then it remains to calculate $t(P_i\otimes P_j)t^{\dagger}$ with $t$ as the corresponding tensor of the node. The results are summarized in Table.~\ref{tab:output_SU2}, where we introduce the symbols $c_1=\cos(2\theta_1)$, $c_2=\cos(2\theta_2)$, $c_{+}=\cos(2\theta_1+2\theta_2)$ and $c_{-}=\cos(2\theta_1-2\theta_2)$ to simplify the notation. For each possible pair of inputs, $\{s,t\}\times \{s,t\}$, we give the results for all possibilities of measurements in the node (no measurement, measurement with singlet outcome, measurement with triplet outcome). It is easy to see that the table is invariant under the operations
\begin{align}
    &(\theta_1,\theta_2)\to(\theta_1+a_1\pi,\theta_2+a_2\pi)\quad a_1,a_2\in \mathbb{Z},\nonumber\\
    &(\theta_1,\theta_2)\to (\theta_2,\theta_1)\nonumber\\
    &(\theta_1,\theta_2)\to (\pi-\theta_2,\pi-\theta_1).
\end{align}
These define the symmetry of the phase diagram. However, it needs to be pointed out that this table only gives the output operator based on the input and measurement outcomes. To correctly capture the recursive relation of probability in Eq.~\eqref{eq:prob_recur_SU2}, we need to multiply elements in the table with an extra factor $c_p=4(1-p)^{(2-\delta)}p^{\delta}$. Here $\delta$ is the number of measurements within the node. Combining Table.~\ref{tab:output_SU2} and this extra factor, we get the coefficients in Eq.~\eqref{eq:sigma_tau_recur} analytically. 

As mentioned in Sec.~\ref{sec:spin_sharpening}, the recursive relation of $\sigma$ and $\tau$ shows that when $p\ne 1$, there can be no sharp phase. When $p=1$, conservation of spin requires that in a shapr phase, $1/4$ fraction of the trajectories must give the singlet state and $3/4$ of the trajectories must give the triplet state. It is worth checking whether this gives a stable fixed point allowed by the recursive relation. 
Considering the distribution with $1/4$ probability to have $\sigma=1$ and $3/4$ probability to have $\tau=1$, Table.~\ref{tab:output_SU2} combined with the value of $c_p$ guarantees that the output distribution is the same as the input distribution. This confirms that the sharp phase we find is an allowed fixed point at $p=1$. It remainst to establish if/when the dynamics flows to this fixed point. We address this problem in Appendix~\ref{append:SU2_lin_recur}.

\section{Theoretical approach to the spin-sharpening transition at $p=1$}
\label{append:SU2_lin_recur}
Here we develop a theoretical approach to the phase diagram of the $SU(2)$-symmetric quantum tree at $p=1$. Before talking about the general case, we first point out two useful facts from Table~\ref{tab:output_SU2}. 
\begin{fact}\label{fact:1}
   When the measurement outcome pair is $(s,s)$, $(s,t)$ or $(t,s)$, the output is always a determined sharp state, no matter what input states are.
\end{fact}
  This is a useful fact which greatly simplifies the theoretical analysis. Based on Fact.~\ref{fact:1}, we further find
  \begin{fact}\label{fact:2}
     When $\theta_1-\theta_2=a\pi$ or $\theta_1+\theta_2=a\pi$ with $a\in \mathbb{Z}$, the dynamics becomes sharp immediately. 
  \end{fact}
  These facts prove the existence of a sharp phase at $p=1$ immediately. In fact, Fig.~\ref{fig:SU2_fixgate} is just an example of Fact~\ref{fact:2}.

Besides the trivial cases in Fact~\ref{fact:2}, the output is not sharpened immediately when the measurement outcome is $(t,t)$. This requires an analytical understanding about the dynamics at late time. To do so, we introduce $\mathcal{Z}$ to represent $\min(\tilde{\sigma},\tilde{\tau})$ with $\tilde{\sigma} = \sigma / (\sigma+\tau)$ and $\tilde{\tau} = \tau / (\sigma + \tau)$. $\mathcal{Z}$ characterizes the spin fluctuation in the density matrix. At the same time, we also need to keep track of which one of between $\tilde{\sigma}$ and $\tilde{\tau}$ is smaller. We classify the density matrices of different trajectories into two sets, the singlet-like trajectories with $\tilde{\tau}=\mathcal{Z}$ and the triplet-like trajectories with $\tilde{\sigma}=\mathcal{Z}$. By introducing an extra bit $\eta=s,t$,
we can label each density matrix by the pair $(\mathcal{Z},\eta)$. Now the singlet-like density matrix is labeled as $(\mathcal{Z},s)$ and the the triplet-like density matrix is labeled as $(\mathcal{Z},t)$. 

Since the trajectories are generally classified into the singlet-like and triplet-like sets, we further define two generating functions
\begin{equation}
    G^s_k(x)=\langle\exp(-e^{-x}\mathcal{Z}_k)\rangle_s\quad G^t_k(x)=\langle\exp(-e^{-x}\mathcal{Z}_k)\rangle_t,
\end{equation}
to describe the two sets separately.
Here the average is taken only over singlet-like/triplet-like trajectories respectively. The wavefronts of the generating functions are located at $\ln\mathcal{Z}_k^{s,typ}=\langle \ln \mathcal{Z}_k\rangle_{s,\mathcal{Z}_k\ne0}$ and $\ln\mathcal{Z}_k^{t,typ}=\langle \ln \mathcal{Z}_k\rangle_{t,\mathcal{Z}_k\ne 0}$ respectively. The sharp phase requires that both $\ln \mathcal{Z}_k^{s,typ}$ and $\ln\mathcal{Z}_k^{t,typ}$decay to $-\infty$ with increase of $k$.

The full recursive relations between $G^{s(t)}_k(x)$ are hard to solve. However, as mentioned in Ref.~\cite{nahum_measurement_2021,feng_measurement_2023}, in the limit where $\mathcal{Z}_k^{typ}$ goes to zero, the non-linear terms in the recursive relation of $\mathcal{Z}$ give an exponentially smaller contribution to the dynamics of the wavefront.
This allows us to consider only the linearized recursive relation of $\mathcal{Z}$. We expect this property is still valid in the dynamics of $SU(2)$ tree. 

One thing which is special about the $SU(2)$ tree is that there is no simple result about the probabilities of $\eta=s$ and $\eta=t$: the type $\eta$ of the output density matrix generally has some complicated dependence on the input density matrices when the measurement outcome pair is $(t,t)$. However, it turns out that when we keep only the linear terms in the recursive relation of $\mathcal{Z}$, the output type $\eta$ becomes independent of $\mathcal{Z}$. 

A possible breakdown of this linearized approximation to the recursive relations comes from trajectories with finite $\mathcal{Z}$ such that the nonlinear term of $\mathcal{Z}$ cannot be ignored. When we are near the wavefront $x=\ln \mathcal{Z}_k^{s(t),typ}$, the effect of trajectories with finite $\mathcal{Z}_k$ is on the order of $\exp(-\mathcal{Z}_k/\mathcal{Z}_k^{s(t),typ})$. Thus the effects of these trajectories are suppressed exponentially when $\mathcal{Z}_k^{s(t),typ}\to 0$. Following this argument, we expect that the dynamics of $G^{s(t)}_k(x)$ in the sharp phase can be solved by ignoring the effects of nonlinear terms of $\mathcal{Z}$ in the recursive relation of $\eta$. 

In Table.~\ref{tab:Z_lead_SU2}, we give output pair $(\mathcal{Z},\eta)$ by ignoring all nonlinear terms of $\mathcal{Z}$ in the recursive relations. The top row of the table gives the types of the input density matrices, while the leftmost column represents the measurement outcomes inside the node\footnote{We slightly abuse the notation $s$ and $t$ to label both the measurement outcome and type $\eta$ (singlet-like or triplet like) of input density matrices.}. At the same time, it is necessary to know the factor of $4p(\tilde{\mathbf{m}}|\rho_1,\rho_2)$ to correctly capture the distribution of measurement outcomes. Using Fact.~\ref{fact:1}, the only nontrivial outcome pair in the recursive relations of $G^s_k(x)$ and $G^t_k(x)$ is $(t,t)$. We have
\begin{equation}
    4p(t,t|\rho_1,\rho_2)\approx 
    \left\{ 
    \begin{aligned} 
    &\frac{9}{8}(1-c_{+})+O(\mathcal{Z}),\quad & \text{[input $(s,s)$]}\\
    &\frac{1}{2}(1-c_{-})+O(\mathcal{Z}),\quad & \text{[input $(s,t)$]}\\
    &\frac{1}{2}(1-c_{-})+O(\mathcal{Z})\quad & \text{[input $(t,s)$]}\\
    &\frac{1}{24}(1-c_{+})+O(\mathcal{Z})\quad & \text{[input $(t,t)$]}.
    \end{aligned}\right.
    \label{eq:SU2_prob_const}
\end{equation} 
 
Using the approximated recursive relation in Table.~\ref{tab:Z_lead_SU2} and Eq.~\eqref{eq:SU2_prob_const}, we get
\begin{widetext}
\begin{align}
G_{k+1}^s(x) &= (\frac{5}{32}+\frac{3}{32}c_{+})+\frac{9}{8}(1-c_{+})G_k^s(x-\ln A_1)G_k^s(x-\ln A_1)+\frac{1}{24}(1-c_{+})G_k^t(x-\ln A_2)G_k^t(x-\ln A_2)\nonumber\\
G^t_{k+1}(x) &= (\frac{9}{16}+\frac{3}{16}c_{+})+(1-c_{-})G_k^s(x+\ln A_2)G_k^t(x+\ln A_1).\label{eq:generation_recur_SU2}
\end{align}
\end{widetext}
Here the constant terms are from the probabilities to get measurement outcome $(s,s)$, $(s,t)$ and $(t,s)$, and to simplify the notation we have introduced the coefficients
\begin{equation}
    A_1 = \frac{4}{9}\frac{1-c_{-}}{1-c_{+}}\quad A_2 =12\frac{1-c_{-}}{1-c_{+}}.
\end{equation}
We see that the recursive relations of $G^s_k(x)$ and $G^t_k(x)$ are coupled with each other. For an exact sharp phase, we have $G_{k\to\infty}^s(x)=1/4$ and $G_{k\to\infty}^t(x)=3/4$. We try the heuristic solutions
\begin{align}
    G_k^{s,\lambda^s}(x)&\sim \frac{1}{4}-\alpha^s_k(\lambda^s)e^{-\lambda^s(x-v^sk)}\nonumber\\
    G_k^{t,\lambda^t}(x)&\sim \frac{3}{4}-\alpha_k^t(\lambda^t)e^{-\lambda^t(x-v^t k)}.\label{eq:attempt}
\end{align}
with $x-vk\gg 1$ and $k\gg 1$ in the sharp phase. Plugging Eq.~\eqref{eq:attempt} into Eq.~\eqref{eq:generation_recur_SU2}, we see that $\lambda^s=\lambda^t$ and $v^s=v^t$ by making $x\to +\infty$ and $t\to +\infty$. This allows us to ignore the superscript of type. Considering the leading term of $e^{-\lambda(x-vk)}$, we get the following condition on $\alpha^s$ and $\alpha^t$:
\begin{align}
\begin{pmatrix}
    \alpha^s\\\\
    \alpha^t
\end{pmatrix}e^{\lambda v}
=
\begin{pmatrix}
    \frac{1}{4}\left(\frac{9}{4}a\right)^{1-\lambda}b^{\lambda}&\frac{3}{4}\left(\frac{1}{12}a\right)^{1-\lambda}b^{\lambda}\\\\
    \frac{3}{4}b^{1-\lambda}\left(\frac{1}{12}a\right)^{\lambda}&\frac{1}{4}b^{1-\lambda}\left(\frac{9}{4}a\right)^{\lambda}
\end{pmatrix}
\begin{pmatrix}
    \alpha^s\\\\
    \alpha^t
\end{pmatrix},
\end{align}
with
\begin{equation}
    a = 1-c_{+},\quad b=1-c_{-}.
\end{equation}

Solving the eigenvalue equation yields
\begin{equation}
    e^{\lambda v}=\frac{B\pm \sqrt{B^2-4C}}{2},
\end{equation}
with 
\begin{equation}
    B=\frac{1}{4}\left[(\frac{9}{4}a)^{1-\lambda}b^{\lambda}+b^{1-\lambda}(\frac{9}{4}a)^{\lambda}\right],\quad C=\frac{3}{32}ab.
\end{equation}
Now there are two possible solutions for each value of $\lambda$, and we need to determine both $\lambda$ and the correct solution. To do so, we take the derivative of the velocity over $\lambda$. At the critical point with $v=0$, we have
\begin{equation}
    \left.\left[\frac{\partial B}{\partial \lambda}\pm \frac{B}{\sqrt{B^2-4C}}\frac{\partial B}{\partial\lambda}\right]\right|_{\lambda^{\ast}}=0
\end{equation}
From $\partial_\lambda B=0$, we get\footnote{The second derivative over $\lambda$ guarantees that $\lambda=1/2$ is the only minimum.}
\begin{equation}
\lambda^{\ast}=1/2.
\end{equation}
Thus at the critical point, $\lambda = 1/2$ and the velocity satisfies
\begin{equation}
    e^{ v/2}=\frac{3\pm\sqrt{3}}{8}\sqrt{ab}.
\end{equation}
It can be seen that if the velocity takes the value $(3-\sqrt{3})\sqrt{ab}/8$, we have $\alpha^s+\alpha^t=0$. But this violates the definition of generating function. So the velocity takes the form
\begin{equation}
    v = 2\ln\left[(\frac{3}{4}+\frac{\sqrt{3}}{4})\left|\sin(\theta_1+\theta_2)\right|\left|\sin(\theta_1-\theta_2)\right|\right].\label{eq:SU2_velocity}
\end{equation}
Then the critical boundary separating sharp and fuzzy phases is given by $v=0$, i.e.
\begin{equation}
\frac{3+\sqrt 3}{4} \left|\sin(\theta_1+\theta_2)\right|\left|\sin(\theta_1-\theta_2)\right| =1.
\end{equation}

\section{Scaling exponents of the spin-sharpening transition at $p=1$}
\label{append:SU2_scale}

In the dynamics of quantum tree obeying $U(1)$ symmetry, we have argued that the purification transition with $d=1$ and the charge-sharpening transition with $d\to+\infty$ both belong to the glass universality class defined in previous work \cite{nahum_measurement_2021,feng_measurement_2023}. Our derivation of the spin-sharpening transition at $p=1$, Appendix~\ref{append:SU2_lin_recur}, also gives $\lambda=1/2$. This suggests that the scaling behavior may be similar. In this section, we give an argument about the scaling behaviors when tuning gate parameters $(\theta_1,\theta_2)$ near the critical boundary at $p=1$. 

We first study the scaling behaviors of order parameter $\mathcal{Z}_{k\to\infty}^{s,typ}$ and $\mathcal{Z}_{k\to\infty}^{t,typ}$ in the fuzzy phase. After long enough time, the wavefronts saturate so we define $\mathcal{Z}^{s(t),typ}=\mathcal{Z}_{k\to\infty}^{s(t),typ}$ in the rest discussion of this section. Considering a point $(\theta_1,\theta_2)$ which is close to the critical boundary $\mathcal{C}$ determined by Eq.~\eqref{eq:SU2_critical_contour}, we should have a saturated generating function $G$ at large enough $k$. In principle, we need to solve the full recursive relation of $G^s_k(x)$ and $G^t_k(x)$ without any approximation. However, as discussed in Refs.~\cite{nahum_measurement_2021,feng_measurement_2023}, when $(\theta_1,\theta_2)$ is quite close to the critical boundary, we can expect that $\min(|\ln \mathcal{Z}^{s,typ}|,|\ln\mathcal{Z}^{t,typ}|)\gg 1$ \footnote{$\ln\mathcal{Z}^{s,typ}$ and $\ln\mathcal{Z}^{t,typ}$ should just have a difference of order $O(1)$ when we are close to the critical boundary. So we can actually ignore their difference in the discussion.}. When $|x|$ is much smaller than the wavefront, we can ignore the nonlinear terms of $G$ in the recursive relations. At the same time, when $x<0$ and $|x|\ll1$, we can ignore the nonlinear terms of $\mathcal{Z}$. Then there is a large regime with $x<0$ and $1\ll |x|\ll \min(|\ln \mathcal{Z}^{s,typ}|,|\ln\mathcal{Z}^{t,typ}|)$ such that inside this regime we can still expect our linearized recursive relations to work accurately. That is, we have
\begin{align}
G^s(x) &= (\frac{5}{32}+\frac{3}{32}c_{+})\nonumber\\
&+\frac{9}{8}(1-c_{+})G^s(x-\ln A_1)G^s(x-\ln A_1)\nonumber
\\
&+\frac{1}{24}(1-c_{+})G^t(x-\ln A_2)G^t(x-\ln A_2)\nonumber\\
G^t(x) &= (\frac{9}{16}+\frac{3}{16}c_{+})\nonumber\\
&+(1-c_{-})G^s(x+\ln A_2)G^t(x+\ln A_1).
\end{align}
Here $G(x)=\lim_{k\to\infty}G_k(x)$ with $1\ll |x|\ll\min(|\ln \mathcal{Z}^{s,typ}|,|\ln\mathcal{Z}^{t,typ}|)$. Inside this regime, we consider the ansatz $H^s(x)=1/4-G^s(x)\sim \alpha^s e^{-\lambda x}$ and $H^t(x)=3/4-G^t(x)\sim \alpha^t e^{-\lambda x}$. Then we have
\begin{align}
\begin{pmatrix}
    \alpha^s\\\\
    \alpha^t
\end{pmatrix}
=
\begin{pmatrix}
    \frac{1}{4}\left(\frac{9}{4}a\right)^{1-\lambda}b^{\lambda}&\frac{3}{4}\left(\frac{1}{12}a\right)^{1-\lambda}b^{\lambda}\\\\
    \frac{3}{4}b^{1-\lambda}\left(\frac{1}{12}a\right)^{\lambda}&\frac{1}{4}b^{1-\lambda}\left(\frac{9}{4}a\right)^{\lambda}
\end{pmatrix}
\begin{pmatrix}
    \alpha^s\\\\
    \alpha^t
\end{pmatrix}.
\end{align}
This eigenvalue equation is solved when
\begin{equation}
    B(\lambda,\theta_1,\theta_2)-C(\theta_1,\theta_2)=1,
\end{equation}
with
\begin{equation}
    B=\frac{1}{4}\left[(\frac{9}{4}a)^{1-\lambda}b^{\lambda}+b^{1-\lambda}(\frac{9}{4}a)^{\lambda}\right],\quad C=\frac{3}{32}ab.
\end{equation}
In this case, $\lambda$ becomes complex to make this equation valid \cite{brunet1997shift}. 
To simplify the discussion, we assume that $(\theta_1,\theta_2)$ is a distance $\delta_\theta$ from the critical boundary $\mathcal{C}$, with the nearest critical point being $(\theta_1^{\ast},\theta_2^{\ast}) \in \mathcal{C}$. When $\delta_{\theta}$ is small enough, $\lambda$ can be solved by the expanding $B-C$ around the point $(\lambda^{\ast}=1/2,\theta_1^{\ast},\theta_2^{\ast})$. We get
\begin{equation}
    \lambda=\frac{1}{2}\pm i\kappa\sqrt{\delta_{\theta}},
\end{equation}
with
\begin{equation}
    \kappa = \left.\sqrt{\frac{2\left|\nabla (B-C)\right|}{\partial^2_{\lambda}B}}\right|_{\lambda^{\ast},\theta_1^{\ast},\theta_2^{\ast}}.
\end{equation}
This result gives us 
\begin{equation}
    H^{s(t)}(x)\sim e^{-\frac{1}{2}(x-x^{s(t)}_0)}\sin(\kappa\sqrt{\delta_{\theta}}x+\phi^{s(t)}).
\end{equation}
Here $x^{s(t)}_0$ and $\phi^{s(t)}$ are some undetermined parameters that are set by matching the behavior of $H(x)$ when $x\to\pm\infty$. To do this, we notice that 
\begin{equation}
   \frac{1}{2}+\partial_x \ln H^{s(t)}= \frac{\kappa\sqrt{\delta_{\theta}}}{\tan(\kappa\sqrt{\delta_{\theta}}x+\phi^{s(t)})},
\end{equation}
when $1\ll |x|\ll \min(|\ln Z^{s,typ}|,|\ln\mathcal{Z}^{t,typ}|)$. When $x<0$ and $|x|\gg|\ln \mathcal{Z}^{s(t),typ}|$, it is easy to see that $\partial_x \ln H=0$. When $x\to+\infty$, $G^{s(t)}(x)\sim 1(3)/4-\langle \mathcal{Z}\rangle_{s(t)} e^{-x}$. Thus we have $\partial_x\ln H^{s(t)}=-1$. To match the asymptotic behavior on the two limits, we have $\phi=\pi/2$ and $\kappa\sqrt{\delta_{\theta}}\ln \mathcal{Z}^{s(t),typ}+\pi=0$ \cite{nahum_measurement_2021}. This gives us the scaling form
\begin{equation}
 \mathcal{Z}^{s(t),typ} \sim \exp(-\frac{K}{\sqrt{\delta_{\theta}}})\quad K=\frac{\pi}{\kappa},
\end{equation}
%%%%\matteo{what is $\kappa$? does it need to appear here? can we just say ``with $K$ a constant''?}
when $(\theta_1,\theta_2)$ approaches the critical point $(\theta_1^{\ast},\theta_2^{\ast})\in\mathcal{C}$ from the fuzzy phase side.

When the parameters $\theta_{1,2}$ sit exactly at a point $(\theta_1^{\ast},\theta_2^{\ast})$ on the critical boundary, the ballistic velocity of $\ln \mathcal{Z}^{s(t),typ}_k$ is zero. However, there can still be a sub-ballistic decay toward $-\infty$ \cite{nahum_measurement_2021}. 
Since the wavefront $\ln \mathcal{Z}^{s(t),typ}_k$ changes slowly with $k$, we use the same conjecture in Refs.~\cite{nahum_measurement_2021,feng_measurement_2023} that within the regime $1\ll |x|\ll |\ln \mathcal{Z}^{s(t),typ}|$, the form of $G^{s(t)}_k(x)$ is similar to the saturated solution in the fuzzy phase. Thus we consider the ansatz  $H^{s(t)}_k(x)\sim \alpha^{s(t)}e^{-\lambda_k (x-f_k)}$, with $f_k$ the location of wavefront. Notice that now $\lambda$ has an imaginary part and it depends on $k$. The position of the wavefront is determined by \cite{nahum_measurement_2021}
\begin{equation}
    f_k \sim -\frac{\pi}{\left|\Im\lambda_k\right|}\label{eq:critical_wavefront}.
\end{equation}
Using the linearized recursive relation of $G_k(x)$, we get
\begin{align}
\begin{pmatrix}
    \alpha^s\\\\
    \alpha^t
\end{pmatrix}e^{-\Delta_k(x)}
=
\begin{pmatrix}
    \frac{1}{4}\left(\frac{9}{4}a\right)^{1-\lambda_k}b^{\lambda_k}&\frac{3}{4}\left(\frac{1}{12}a\right)^{1-\lambda_k}b^{\lambda_k}\\\\
    \frac{3}{4}b^{1-\lambda_k}\left(\frac{1}{12}a\right)^{\lambda_k}&\frac{1}{4}b^{1-\lambda_k}\left(\frac{9}{4}a\right)^{\lambda_k}
\end{pmatrix}
\begin{pmatrix}
    \alpha^s\\\\
    \alpha^t
\end{pmatrix},
\end{align}
where we defined $\Delta_k(x)=\lambda_{k+1}x-\lambda_kx -\lambda_{k+1}f_{k+1}+\lambda_kf_k$. 
At large enough $k$, the imaginary part of $\lambda$ should decay to $0$ since the wavefront goes to $-\infty$. At the same time, the speed of $f_k$ also decays to $0$ due to the sub-ballistic motion of the wavefront. Using these facts, we can approximate $\Delta_k$ by\footnote{Notice that since $1\ll|x|\ll\min(|\ln\mathcal{Z}^{s,typ}|,|\ln\mathcal{Z}^{t,typ}|)$, the part of $\Delta_k$ that depends on $x$ can be ignored.} $-(f_{k+1}-f_k)/2$. 
Then we get
\begin{equation}
    \Im\lambda= \left.\frac{\sqrt{2-B}}{\sqrt{\partial^2_{\lambda} B}}\right|_{\lambda=1/2,\mathcal{C}}\sqrt{|f_{k+1}-f_k|}\label{eq:critical_dlambda}.
\end{equation}
Combining Eq.~\eqref{eq:critical_wavefront} and Eq.~\eqref{eq:critical_dlambda}, we get
\begin{equation}
    f_k \sim -k^{1/3},
\end{equation}
and thus
\begin{equation}
    \ln \mathcal{Z}^{s(t),typ}\sim -k^{1/3}
\end{equation}
on the critical contour $\mathcal{C}$.

So far we have argued that the spin-sharpening transition at $p=1$ obtained by tuning the gate parameter $\theta_{1,2}$ may still be in the glass universality class. It would also be interesting to understand the critical behaviors when we fix $\theta_1$ and $\theta_2$ and let $p$ approach $1$. We leave this question in future work.

\begin{table*}
\centering
\begin{adjustbox}{width=0.9\textwidth}
\renewcommand{\arraystretch}{1.3}
\begin{tabular}{|c|c|c|c|c|}
   \hline
   \diagbox{$m's$}{input}&$s,s$&$s,t$& $t,s$& $t,t$\\
   \hline
   $s$,$s$ &$(0,s)$&$(0,s)$&$(0,s)$&$(0,s)$\\
   \hline
   $s$,$t$&$(0,t)$&$(0,t)$&$(0,t)$& $(0,t)$\\
   \hline
   $t$,$s$&$(0,t)$&$(0,t)$&$(0,t)$&$(0,t)$\\
   \hline
   $t$,$t$&$(\frac{4(1-c_{-})}{9(1-c_{+})}(\mathcal{Z}_1+\mathcal{Z}_2),s)$&$(\frac{1-c_{+}}{1-c_{-}}(\frac{\mathcal{Z}_1}{12}+\frac{4\mathcal{Z}_2}{9}),t)$&$(\frac{1-c_{+}}{1-c_{-}}(\frac{4\mathcal{Z}_1}{9}+\frac{\mathcal{Z}_2}{12}),t)$&$(12\frac{1-c_{-}}{1-c_{+}}(\mathcal{Z}_1+\mathcal{Z}_2),s)$\\
   \hline
\end{tabular}
\end{adjustbox}
\caption{Values of the node output parameters $(\mathcal{Z},\eta)$, to linear order in $\mathcal{Z}_{1,2}$, at $p=1$. The top row labels the types (singlet $s$ or triplet $t$) of the two input density matrices to the node, while the leftmost column labels the two mid-curcuit measurement outcomes (also singlet $s$ or triplet $t$), where the first outcome corresponds to the inner qubit pair and the second corresponds to the outer qubit pair in Fig.~\ref{fig:SU2_tree}. We again use the shorthand notation $c_i = \cos(2\theta_i)$, $i=1,2$, and $c_\pm = \cos(2(\theta_1\pm \theta_2))$. Here we assume that $\theta_1+\theta_2 \notin \pi\mathbb{Z}$ and $\theta_1-\theta_2\notin \pi\mathbb{Z}$ (otherwise the system sharpens immediately).}
\label{tab:Z_lead_SU2}
\end{table*}

\section{Alternative measurement outcome distributions}
\label{append:fake_measurement}

While real quantum measurements are distributed according to the Born rule, it is sometimes valuable to consider alternative (unphysical) distributions that may be more analytically or numerically tractable, while still shedding light on certain features of the ensemble of quantum trajectories~\cite{Bao_Theory_2020,Bao_2021}.

One such distributions is given by {\it forced measurements}~\cite{nahum_measurement_2021}, where an output string is chosen uniformly at random (among allowed outcomes, i.e. those that have $p({\mathbf m})>0$) and corresponding projectors are then applied to the circuit. This corresponds to replacing $p({\mathbf m}) \mapsto {\rm const}>0$ for all $\mathbf m$ with $p(\mathbf m)>0$. 
More generally, one can study the distribution $p({\mathbf m}) \mapsto p(\mathbf m)^n / \sum_{\mathbf m'} p(\mathbf m')^n$, for any value of $n$; $n=0$ gives forced measurements, $n=1$ gives the Born rule, and $n>1$ gives modified distributions that concentrate more weight to the more likely trajectories in the Born distribution. $n$ plays the role of a ``replica number'' in statistical-mechanical treatments of the MIPT. 

We can study spin sharpening in these modified trajectory ensembles by introducing the family of order parameters
\begin{align}     
r_n & =\sum_{\mathbf{m}}\frac{(\sigma(\mathbf{m})+\tau(\mathbf{m}))^n}{\sum_{\mathbf{m}' }(\sigma(\mathbf{m}')+\tau(\mathbf{m}'))^n}\frac{\sigma(\mathbf{m})^2+\tau(\mathbf{m})^2}{(\sigma(\mathbf{m})+\tau(\mathbf{m}))^2} \nonumber \\
& = \frac{\sum_{\mathbf{m}}(\sigma(\mathbf{m})+\tau(\mathbf{m}))^{n-2}(\sigma(\mathbf{m})^2+\tau(\mathbf{m})^2)}{\sum_{\mathbf{m}}(\sigma(\mathbf{m})+\tau(\mathbf{m}))^n}
\label{eq:fake_order}.
\end{align}

In Eq.~\eqref{eq:fake_order}, we are particularly interested in the cases of $n=0$ and $n\ge 2$. When $n=0$, it means that all trajectories have the same probability as long as $\sigma+\tau\ne 0$. This corresponds to all mid-circuit measurements being forced, a case that can be efficiently simulated by the pool method (see Refs.~\cite{nahum_measurement_2021,feng_measurement_2023}). 

Another tractable case is when $n\ge 2$ and $n$ is integer. Then, Eq.~\eqref{eq:fake_order} can be expressed purely in terms of the variables 
\begin{equation}
    x_l \equiv \sum_{\mathbf m} \sigma(\mathbf m)^l \tau(\mathbf m)^{n-l},
\end{equation}
for $0\leq l \leq n$. Specifically, we have
\begin{equation}
    r_n = \frac{\sum_{l = 0}^{n-2} \binom{n-2}{l} (x_l+x_{n-l}) }{\sum_{l=0}^n \binom{n}{l} x_l}.
\end{equation}
Using the recursive relations in Eq.~\eqref{eq:sigma_tau_recur}, we can efficiently express the variables $\{x_l\}_{l=0}^n$ for a $k$-layer tree from those of a $(k-1)$-layer tree with a quadratic recursion, and thus get $r_n$ for in principle arbitrarily large values of the tree depth $k$.

\subsubsection{Forced measurements ($n=0$)}
In the tree model with only forced measurement, the dynamics can be efficiently simulated by the pool method as mentioned before.  We find that when all allowed trajectories ($\sigma+\tau\ne 0$) have equal probability, the dynamics flows to a sharp state no matter the choice of $\theta_{1}$ and $\theta_2$. This suggests that within all allowed trajectories, the spin-sharp ones are prevalent; a fuzzy phase must require some over-weighting of the less-common fuzzy trajectories in the ensemble.

\subsubsection{Non-Born-rule measurements ($n\geq 2$)}
\begin{figure}
    \centering
    \includegraphics[width=0.9\columnwidth]{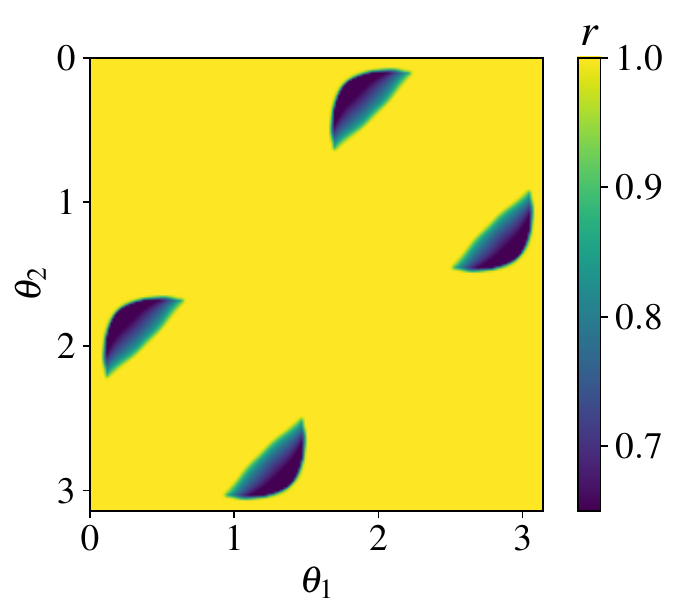}
    \caption{Simulation of $SU(2)$-symmetric tree with $n=2$-replica measurement outcome distribution at $p=1$. We use the efficient recursion method described in the main text and iterate up to a well-converged value of $k=400$.}
    \label{fig:SU2_fake}
\end{figure}

For all integer $n\ge 2$, we find similar behavior up to the numerically accessible $n \approx 10$. This allows us to focus on $n=2$ to illustrate the phenomenology. Our results for the order parameter $r_2$ at very large $k$ are shown in Fig.~\ref{fig:SU2_fake}. 
Here we observe that, although most of the phase diagram is in a sharp phase, there are also very clear fuzzy phases. This behavior holds for all numerically accessible $n$, which suggests that the most likely (i.e., highest-Born-probability) trajectories in the ensemble are fuzzy in these parts of parameter space. 

\begin{figure}
    \centering
    \includegraphics[width=1.0\columnwidth]{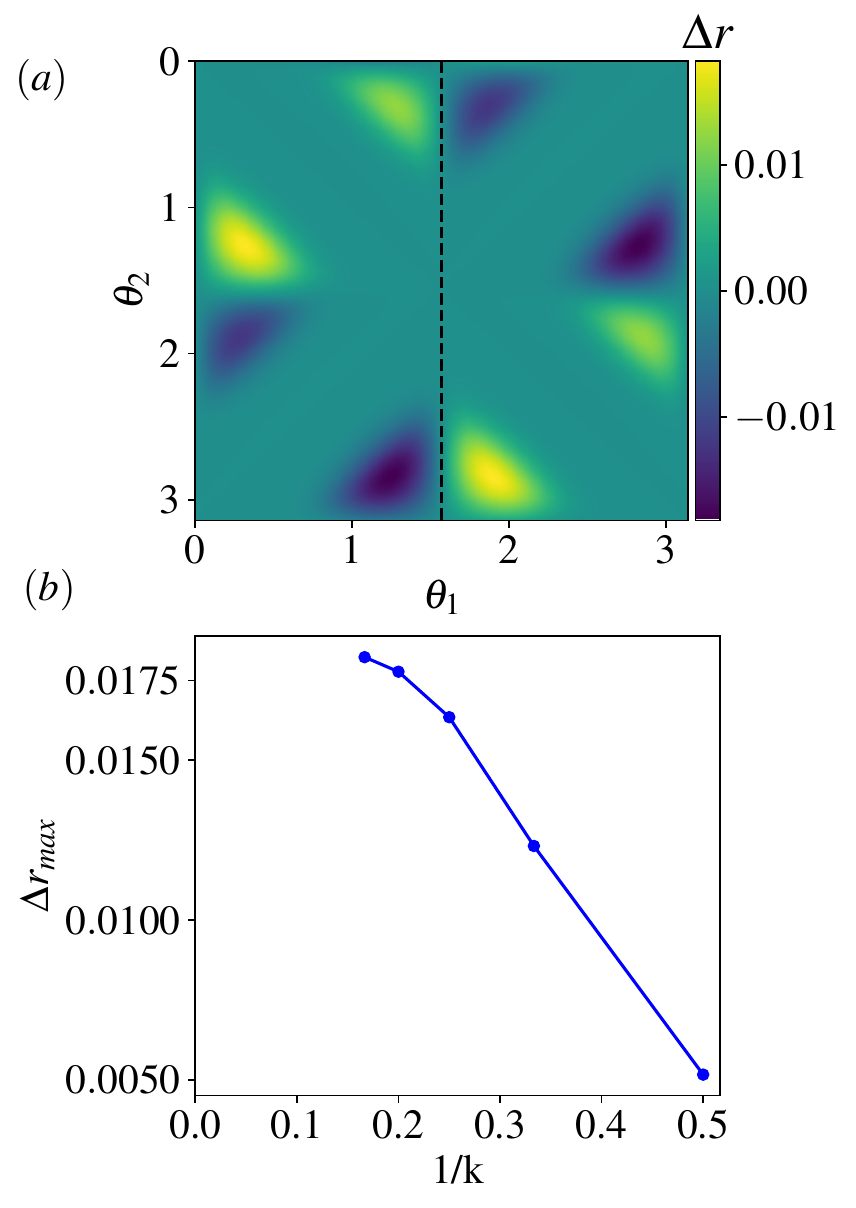}
    \caption{(a) Spin sharpening order parameter $r$ for Born-rule measurement ($n=1$) at $p=1$, from Fig.~\ref{fig:real_128}(a), antisymmetrized about the $\theta_1 = \pi/2$ axis (dashed line): $\Delta r(\theta_1,\theta_2) = r(\pi- \theta_1,\theta_2) - r(\theta_1,\theta_2)$.
    (b) Maximum of $\Delta r$ vs $1/k$ suggests a persistent asymmetry in $r$ in the $k\to\infty$ limit. }
    \label{fig:real_antisym}
\end{figure}

Interestingly, the phase boundaries in this case are strongly asymmetric under the reflection $\theta_i\mapsto \pi-\theta_i$, unlike the symmetric contour we identified for the Born-rule ($n=1$) measurements, Eq.~\eqref{eq:SU2_critical_contour}. 
The existence of this asymmetric pattern in the $n\geq 2$ measurements raises the natural question of whether hints of such a pattern can also be seen in the real measurement case. 
To check this, we antisymmetrize the real-measurement data (obtained for $k=6$, i.e. 128 system qubits) about the $\theta_1 = \pi/2$ axis, defining $\Delta r(\theta_1,\theta_2) = r(\pi-\theta_1,\theta_2) - r(\theta_1,\theta_2)$; 
results are shown in Fig.~\ref{fig:real_antisym}(a). 
Interestingly, this reveals a very similar anti-symmetric pattern as for the $n=2$ measurements. 

We further check the dependence of this asymmetry on the tree depth $k$ in Fig.~\ref{fig:real_antisym}(b), where we plot $\max_{\theta_{1,2}} \Delta r$ as a function of $1/k$. The results, while finite-size limited, are suggestive of a persistent asymmetry in the $k\to\infty$ limit, indicating that this is a real feature of the steady state and is not transient. 

In summary, we have that:
\begin{enumerate}
    \item Almost all trajectories (by number, not by Born probability) are sharp across the phase diagram, as shown by the $n=0$, forced-measurement result; 
    \item The Born-rule average across trajectories becomes fuzzy in roughly circular regions, as shown by the $n=1$ data in Fig.~\ref{fig:real_128} and the analytical phase boundary Eq.~\eqref{eq:SU2_critical_contour};
    \item The most likely (i.e. highest Born rule probability) trajectories become fuzzy in roughly semi-cricular regions, as shown by the $n=2$ data in Fig.~\ref{fig:SU2_fake}; these do {\it not} match the $n=1$ phase boundary: in parts of the sharp phase the most likely trajectories become fuzzy, and vice versa. 
\end{enumerate}
This is evidence of additional interesting structure in the ensemble of trajectories that goes beyond the Born-averaged spin sharpening. We leave further investigations of these structures to future work.

\end{document}